\documentclass[journal]{IEEEtran}

\usepackage[export]{adjustbox}
\usepackage{hyperref}
\hypersetup{final=true}
\usepackage{amsfonts}
\usepackage[cmex10]{amsmath}
\usepackage{amssymb}
\usepackage{color}
\usepackage{latexsym}
\usepackage[table,xcdraw]{xcolor} 
\usepackage{caption}
\usepackage{subcaption}
\usepackage{graphicx}
\usepackage{setspace}
\usepackage{multirow}
\usepackage{cite}
\usepackage{enumerate}
\usepackage{enumitem}
\usepackage{tabularx}
\newtheorem{lemma}{Lemma}

\graphicspath{ {figures/} {results_exp/} {results_sim/} }

\begin{document}

\title{A Practical Spectrum Sharing Scheme for Cognitive Radio Networks: Design and Experiments}

\author{\IEEEauthorblockN{
	Pedram~Kheirkhah~Sangdeh, 
	Hossein~Pirayesh,
	Adnan~Quadri, and
	Huacheng~Zeng
	}	

	\IEEEcompsocitemizethanks{
	\IEEEcompsocthanksitem The authors are with the Department of Electrical and Computer Engineering, University of Louisville, Louisville, KY  40292. \protect  }
	\thanks{Part of this work was presented in IEEE Infocom 2019 \cite{kheir2019practical}.}
}

\maketitle

\begin{abstract}
Spectrum shortage is a fundamental problem in wireless networks and this problem becomes increasingly acute with the rapid proliferation of wireless devices. 
To address this problem, spectrum sharing in the context of cognitive radio networks (CRNs) has been considered a promising solution. 
In this paper, we propose a practical spectrum sharing scheme for a small CRN that comprises a pair of primary users and a pair of secondary users by leveraging the multiple-input and multiple-output (MIMO) technology. 
In our scheme, we assume that the secondary users take full responsibility for cross-network interference cancellation (IC). 
We also assume that the secondary users have no knowledge about the primary network, including its signal waveform, frame structure, and network protocol. 
The key components of our proposed scheme are two MIMO-based interference management techniques: 
blind beamforming (BBF) and blind interference cancellation (BIC).
We have built a prototype of our scheme on a wireless testbed and demonstrated that the prototyped secondary network can coexist with commercial Wi-Fi devices (primary users).
Experimental results further show that, for a secondary device with two or three antennas, BBF and BIC achieve an average of 25~dB and 33~dB IC capability in an office environment, respectively.

\begin{IEEEkeywords}
Spectrum sharing, coexistence, cognitive radio networks, blind interference cancellation, blind beamforming
\end{IEEEkeywords}

\end{abstract}

\section{Introduction}
\label{sec:introduction}

The burgeoning demands for data-hungry wireless services and rapid proliferation of wireless devices (e.g., mobile devices and the Internet-of-Things sensors) have pushed the spectrum shortage issue to a breaking point. 
Although it is expected that much spectrum in the millimeter band (30~GHz to 300~GHz) will be allocated for communication purposes, most of this spectrum might be limited to short-range communications due to the severe path loss. 
Moreover, millimeter band is highly vulnerable to blockage and thus mainly considered for complementary use.
As envisioned, sub-6~GHz frequency spectrum, which has already been very crowded, will still be the main spectrum band to carry the majority of wireless traffic for a long time in commercial wireless systems.
Therefore, it is a crucial problem to maximize the sub-6 GHz spectrum utilization efficiency.

To improve spectrum utilization efficiency, spectrum sharing in the context of cognitive radio networks (CRNs) has been regarded as a promising and cost-effective solution.
In the past two decades, CRNs have received a large amount of research efforts and have produced many results that allow cognitive users (secondary users) to coexist with non-cognitive users (primary users) without causing harmful interference to primary users' communications. 
Depending on the knowledge and communications strategy that are needed at the secondary users, cognitive radio falls into three paradigms: 
interweave, overlay, and underlay \cite{goldsmith2009breaking}.
In the interweave paradigm, the secondary users exploit spectrum white holes and intend to access the spectrum opportunistically when the primary users are idle. 
In the overlay paradigm, the secondary users are allowed to access spectrum simultaneously with the primary users, provided that the primary users share the knowledge of their signal codebooks and messages with the secondary users. 
Compared to these two paradigms, the underlay paradigm is more appealing as it allows secondary users to concurrently utilize the spectrum with primary users while requiring neither cooperation nor knowledge from the primary users.

Despite a large body of work on underlay CRNs, most of the existing work is either focused on theoretical exploration 
or reliant on unrealistic assumptions such as cross-network channel knowledge and inter-network cooperation (see, e.g., 
\cite{lan2017optimal,wang2015optimal,liu2016nonorthogonal,dadallage2016joint,nguyen2017efficient,kusaladharma2017secondary}). 
Very limited progress has been made so far in the development of practical schemes to enable spectrum sharing in underlay CRNs.
To the best of our knowledge, there is no underlay spectrum sharing scheme that has been implemented and evaluated in real-world wireless environments.
The key challenge in the design of practical underlay spectrum sharing schemes lies in the management of cross-network interference, which is reflected in the following two tasks: 
(i) at a secondary transmitter, how to pre-cancel its generated interference for the primary receivers in its vicinity; and 
(ii) at a secondary receiver, how to decode its desired signal in the presence of interference from primary transmitters. 
These two tasks become particularly challenging in the CRNs where the secondary users do not know the signal waveform, frame structure, and network protocol of the primary network and where the primary users are unable or unwilling to cooperate with the secondary users.

In this paper, we propose a practical spectrum sharing scheme for a small CRN that comprises a pair of primary users and a pair of secondary users. 
We assume that the primary users are oblivious to the secondary users, and the secondary users do not know the signal waveform, frame structure, and network protocol of the primary network. 
We also assume that the secondary users have more antennas than the primary users and that the secondary users take the full responsibility for cross-network interference cancellation~(IC). 
Our scheme takes advantage of the recent advances in multiple-input and multiple-output~(MIMO) technology to tame the cross-network interference.
The key components of our scheme are two MIMO-based interference management techniques: 
blind beamforming~(BBF) and blind interference cancellation~(BIC).

The proposed BBF technique is used at the secondary transmitter to avoid introducing interference at the primary receiver. 
In contrast to existing beamforming techniques, which require channel knowledge for the construction of beamforming filters, our BBF technique does not require channel knowledge. 
Instead, it constructs the beamforming filters to avoid interference for a primary user by leveraging the overheard signals from that primary user. 
The proposed BIC technique is used at the secondary receiver to decode its desired signals in the presence of interference from the primary transmitter.
Again, different from existing IC techniques, which require channel state information (CSI) and inter-network synchronization, our BIC technique requires neither the cross-network channel knowledge or inter-network synchronization for signal detection.
Instead, it leverages the reference signal (preamble) embedded in data frame to construct the decoding filters for signal detection in the face of interference. 
Collectively, these two techniques effectively tame the cross-network interference from the secondary network side, without requiring knowledge or coordination from the primary network.

We have built a prototype of our scheme on a wireless testbed to evaluate its feasibility and performance in real-world wireless environments.
As an example, we demonstrate that our prototyped secondary devices can coexist with commercial off-the-shelf (COTS) Wi-Fi devices (primary users).
The secondary users achieve 1.2~bits/s/Hz spectral utilization without harmfully affecting the packet delivery rate of Wi-Fi communications.
A video of our demo is presented in \cite{Demo2019practical}. 
We further conduct experiments to evaluate the performance of the secondary network in coexistence with LTE-like and CDMA-like primary networks in the following two cases: 
(i) the primary users equipped with one antenna and the secondary users equipped with two antennas; and 
(ii) the primary users equipped with two antennas and the secondary users equipped with three antennas.
Experimental results measured at 12 different locations in an office environment show that the secondary network can achieve 1.1~bits/s/Hz spectral utilization without harmfully degrading the performance of primary networks. 
Moreover, experimental results show that the proposed BBF and BIC techniques achieve an average of 25~dB and 33~dB IC capability over the tested 12 locations, respectively.

The remainder of this paper is organized as follows.
Section~\ref{sec:related_work} surveys the related work.  Section~\ref{sec:Problem_statement} clarifies the problem and system model. 
Section~\ref{sec:protocol} offers an overview of the proposed spectrum sharing scheme at the MAC and PHY layers. 
Section~\ref{sec:bbf} and Section\ref{sec:bic} present the proposed BBF and BIC techniques, respectively. 
Section~\ref{sec:evaluation} exhibits our experimental results.
Section~\ref{sec:limitation} discusses the limitations of our scheme, and Section~\ref{sec:conclusion} finally concludes this paper.

\section{Related Work}
\label{sec:related_work}

We focus our literature survey on spectrum sharing in underlay CRNs and the related interference management techniques.

\noindent
\textbf{Spectrum Sharing in Underlay CRNs.}
Underlay CRNs allow simultaneous spectrum utilization in both primary and secondary networks as long as the interference level at the primary users remains acceptable.
Different signal processing techniques have been studied for interference management in underlay CRNs,
such as
spread spectrum \cite{horne2003adaptive}, 
power control \cite{lan2017optimal,wang2015optimal,liu2016nonorthogonal}, 
and beamforming 
\cite{pennanen2014multi,gharavol2010robust,xu2015robust,huang2012robust,alabbasi2014energy,filippou2016coordinated,he2014leakage,he2014sum,afana2014performance,afana2015cooperative,nandan2018maximizing,nandan2018secure,zhang2018secure,zhang2010cognitive,chen2013interference,al2016transmit,cai2014cognitive,noam2013blind,xu2013practical}.
%
%
While spread spectrum handles interference in the spectral domain and power control tames interference in the power domain, the beamforming technique exploits the spatial degrees of freedom~(DoF) provided by multiple antennas to steer the secondary signals to some particular directions, thereby avoiding interference for primary users.
Compared to the other two techniques, beamforming is more appealing in practice as it is more effective in interference management.

Given its potential, beamforming has been studied in underlay CRNs to pursue various objectives, such as 
improving energy efficiency of secondary transmissions
\cite{pennanen2014multi,gharavol2010robust,xu2015robust,huang2012robust}, 
maximizing data rate of secondary users \cite{afana2014performance,afana2015cooperative}, 
maximizing sum rate of both primary and secondary users
\cite{alabbasi2014energy,filippou2016coordinated,he2014leakage,he2014sum}, 
and enhancing the security against eavesdroppers \cite{nandan2018maximizing,nandan2018secure,zhang2018secure}.   
%
%
%
%
However, most of these beamforming solutions are reliant on global network knowledge and cross-network channel knowledge. 
Our work differs from these efforts as it requires neither cross-network channel knowledge nor inter-network cooperation.

\noindent
\textbf{BBF in Underlay CRNs.} 
There is some pioneering work that studied BBF to eliminate the requirement of cross-network channel knowledge for the design of beamforming filters \cite{zhang2010cognitive,al2016transmit,cai2014cognitive,chen2013interference,noam2013blind,xu2013practical}.
In \cite{zhang2010cognitive} and \cite{chen2013interference}, an eigen-value-decomposition-based approach was proposed to construct beamforming filters at the secondary transmitter using its received interfering signals from the primary device. 
When the secondary device transmitting, the constructed beamforming filters would steer its radio signals to the null subspace of the cross-network channel, thereby avoiding interference for the primary device.
Our BBF technique follows similar idea, but differs in the network setting and design objective.
Specifically, these two efforts conducted theoretical analysis to optimize the data rate of secondary users under certain interference temperature, while the BBF technique in our work is developed with joint consideration of its practicality and performance in real OFDM-based networks.

In \cite{al2016transmit} and \cite{cai2014cognitive}, the beamforming design is formulated as a part of a network optimization problem, and some constraints are developed based on statistical channel knowledge to relax the requirement of cross-network channel knowledge.  
This approach is of high complexity, and it seems not amenable to practical implementation. 
In \cite{noam2013blind} and \cite{xu2013practical}, spatial learning methods were proposed to iteratively adjust beamforming filters at the secondary devices based on the power level of the primary transmission, with the objective of reducing cross-network interference for the primary users.
These methods are cumbersome and not amenable to practical use.

%
%
%
%
%
%
%
%
%
%

\noindent
\textbf{MIMO-based BIC.} 
While there are many results on interference cancellation in cooperative wireless networks, the results of MIMO-based BIC in non-cooperative networks remain limited. 
In~\cite{rousseaux2002blind}, Rousseaux et al. proposed a MIMO-based BIC technique to handle interference from one source.
In~\cite{winters1993signal}, Winters proposed a spatial filter design for signal detection at multi-antenna wireless receivers to combat unknown interference. 
In~\cite{gollakota2011clearing}, Gollakota et al. proposed a MIMO-based solution to mitigate narrow-band interference from home devices such as microwave.
Furthermore, BIC has been studied in the context of radio jamming in wireless communications (see, e.g., \cite{yan2016jamming,shen2014mcr}). 
%
%
Compared to the existing BIC techniques, our BIC technique has a lower complexity and far better performance (33~dB IC capability in our experiments).

\section{Problem Statement}
\label{sec:Problem_statement}
\begin{figure}
	\includegraphics[width=3.5in]{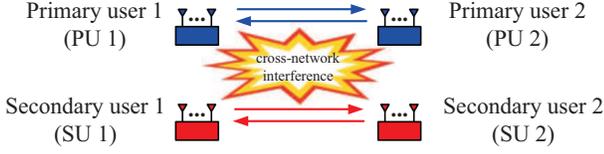}
	\caption{A CRN consisting of two active primary users and two active secondary users.}
	\label{fig:crn}
\end{figure}

\begin{figure}
	\centering
	\includegraphics[width=3.5in]{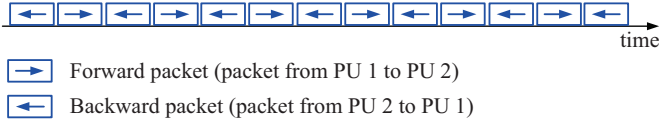}
	\caption{Consistent and persistent traffic in the primary network.}
	\label{fig:primary_user_traffic_pattern}
\end{figure}

We consider an underlay CRN as shown in Fig.~\ref{fig:crn}, which consists of two active primary users and two secondary users. 
The primary users establish bi-directional communications in time-division duplex (TDD) mode. 
The traffic flow in the primary network is persistent and consistent in both directions, as shown in Fig.~\ref{fig:primary_user_traffic_pattern}.
The secondary users want to utilize the same spectrum for their own communications. 
To do so, the secondary transmitter employs beamforming to pre-cancel its generated interference for the primary receiver; and the secondary receiver performs IC for its signal detection. 
Simply put, the secondary users take full burden of cross-network interference cancellation, and their data transmissions are transparent to the primary users.

In this CRN, there is no coordination between the primary and secondary users. 
The secondary users have no knowledge about cross-network interference characteristics.
The primary users have one or multiple antennas, and the number of their antennas is denoted by $M_\mathrm{p}$.
The secondary users have multiple antennas, and the number of their antennas is denoted by $M_\mathrm{s}$.
In our study, we assume that the number of antennas on a secondary user is greater than that of a primary user (i.e., $M_\mathrm{s} > M_\mathrm{p}$).
This assumption ensures that each secondary user has sufficient spatial DoF to tame the cross-network interference from/to the primary users. 

\noindent
\textbf{Our Objective.} 
In such a CRN, our objective is four-fold:
(i)~develop a BBF technique for the secondary transmitter to pre-cancel its generated interference at the primary receiver;
(ii)~develop a BIC technique for the secondary receiver to decode its desired signals in the presence of interference from the primary transmitter;
(iii)~design a spectrum sharing scheme by integrating these two IC techniques;
and 
(iv)~evaluate the IC techniques and the holistic scheme by experimentation.

\noindent
\textbf{Two Justifications.}
First, in this paper, we study a CRN that comprises one pair of primary users and one pair of secondary users.
Such a CRN, albeit small in the network size, serves as a fundamental building block for a large-scale CRN that have many primary and secondary users. 
Therefore, understanding this small CRN is of both theoretical and practical importance. 
Second, in our study, we assume that the secondary users have no knowledge about cross-network interference characteristics.
Such a conservative assumption leads to a more robust spectrum sharing solution, which is suited for many application scenarios.

\section{A Spectrum Sharing Scheme}
\label{sec:protocol}	
\begin{figure}
	\centering
	\includegraphics[width=3.5in]{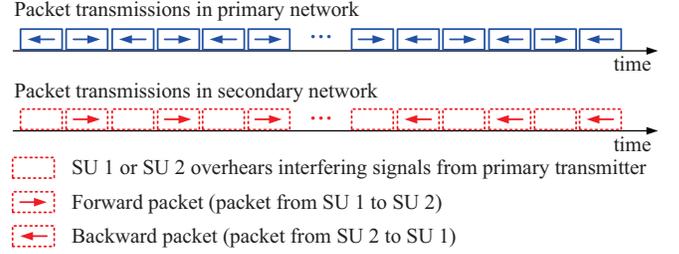}
	\caption{A MAC protocol for spectrum sharing in a CRN that has two primary users and two secondary users.}
	\label{fig:mac_protocol}
\end{figure}

\begin{figure}
	\centering
	\begin{subfigure}[b]{1.65in}
		\includegraphics[width=1.7in]{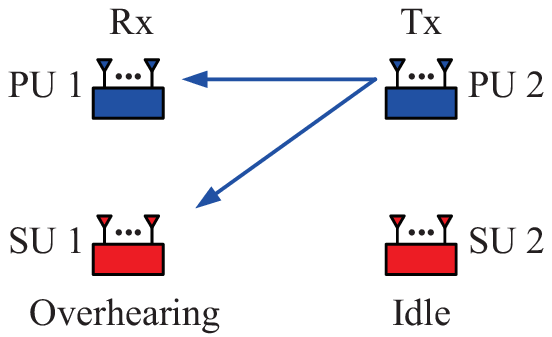}
		\caption{Phase I: SU 1 overhears the interfering signals from PU 2.}
	\end{subfigure}
	~~
	\begin{subfigure}[b]{1.65in}
		\includegraphics[width=1.7in]{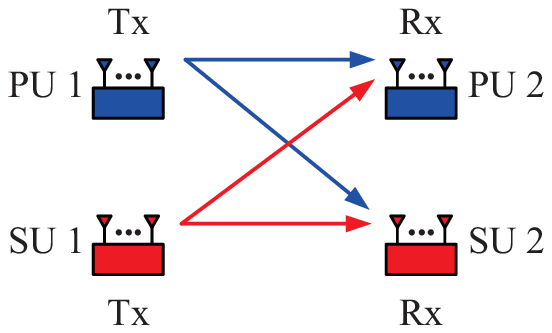}
		\caption{Phase II: SU 1 sends data to SU 2 using IC techniques.}
	\end{subfigure}
	\caption{Illustration of our proposed spectrum sharing scheme.}
	\label{fig:spectrum_sharing_scheme}
\end{figure}

In this section, we present a spectrum sharing scheme for the secondary network so that it can use the same spectrum for its communications without adversely affecting the performance of the primary network. 
Our scheme consists of a lightweight MAC protocol and a new PHY design for the secondary users.
In what follows, we first present the MAC protocol and then describe the new PHY design.

\subsection{MAC Protocol for Secondary Network}
Fig.~\ref{fig:mac_protocol} shows our MAC protocol in the time domain.
It includes both forward communications (from SU 1 to SU 2) and backward communications (from SU 2 to SU 1) between the two secondary users. 
Since the two communications are symmetric, our presentation in the following will focus on the forward communications. 
The backward communications can be done in the same way.

The forward communications in the proposed MAC protocol comprise two phases: \textit{overhearing (Phase I)} and \textit{packet transmission (Phase II)}.
In the time domain, Phase I aligns with the backward packet transmissions of the primary network, and Phase II aligns with the forward packet transmissions of the primary network, as illustrated in Fig.~\ref{fig:mac_protocol}.
We elaborate the operations in the two phases as follows:

\begin{itemize}
\item  
\textbf{Phase~I:} 
SU~1 overhears the interfering signals from PU~2 and SU~2 remains idle, as shown in Fig.~\ref{fig:spectrum_sharing_scheme}(a).

\item 
\textbf{Phase~II:}
SU~1 first constructs beamforming filters using the overheard interfering signals in Phase~I and then transmits signals to SU~2 using the constructed beamforming filters.
Meanwhile, SU~2 decodes the signals from SU~1 in the presence of interference from PU~1.
Fig.~\ref{fig:spectrum_sharing_scheme}(b) shows packet transmissions in this phase.
\end{itemize}

\subsection{PHY Design for Secondary Users: An Overview}

To support the proposed MAC protocol, we use the IEEE~802.11 legacy PHY for the secondary network, including the frame structure, OFDM modulation, and channel coding schemes.
However, IEEE 802.11 legacy PHY is vulnerable to cross-network interference.
Therefore, we need to modify the legacy PHY for the secondary users.
The modified PHY should be resilient to cross-network interference on both transmitter and receiver sides.
The design of such a PHY faces the following two challenges.

\noindent
\textbf{Challenge 1.}
Referring to Fig.~\ref{fig:spectrum_sharing_scheme}(b), the main task of the secondary transmitter (SU~1) is to pre-cancel its generated interference at the primary receiver (PU~2). 
Note that we assume the secondary transmitter has no knowledge about the primary network, including the signal waveform, bandwidth, and frame structure.
The primary network may use OFDM, CDMA, or other types of modulation for packet transmission. 
The lack of knowledge about the interfering signals from the primary transmitter makes it challenging to manage the interference on the secondary network. 

To address this challenge, we design a BBF technique for the secondary transmitter (SU~1) to pre-cancel its interference at the primary receiver.
Our beamforming technique takes advantage of the overheard interfering signals in Phase~I to construct precoding vectors for beamforming.
Our BBF technique, albeit without knowledge of interfering signals, can completely pre-cancel the interference at the primary receiver if the noise is zero and the forward/backward channels are reciprocal. 
Details of this beamforming technique are presented in Section~\ref{sec:bbf}.

\noindent
\textbf{Challenge 2.}
Again, referring to Fig.~\ref{fig:spectrum_sharing_scheme}(b), the main task of the secondary receiver (SU 2) is to decode its desired signals in the presence of cross-network interference from the primary transmitter.
Note that the secondary receiver has no knowledge of the interference characteristics.
This makes it challenging to cancel interference for signal detection. 

To address this challenge, we design a MIMO-based BIC technique for the secondary receiver.
The core component of our BIC technique is a spatial filter, which mitigates the (unknown) cross-network interference from the primary transmitter and recovers the desired signals.
Details of this BIC technique are presented in Section~\ref{sec:bic}.

\section{Blind Beamforming}
\label{sec:bbf}	

In this section, we study the beamforming technique at SU~1 in Fig.~\ref{fig:spectrum_sharing_scheme}.
In Phase~I, SU~1 first overhears the interfering signals from the primary transmitter, and then uses the overheard interfering signals to construct spatial filters.
Based on channel reciprocity, the constructed spatial filters are used as beamforming filters in Phase~II to avoid interference at the primary receiver. 
These operations are performed on each subcarrier in the OFDM modulation. 
In what follows, we first present the derivation of beamforming filters and then offer performance analysis of the proposed beamforming technique.

\noindent
\textbf{Mathematical Formulation.}
Consider SU~1 in Fig.~\ref{fig:spectrum_sharing_scheme}(a). 
It overhears interfering signals from PU~2. 
The overheard interfering signals are converted to the frequency domain through FFT operation.\footnote{The interfering signals are not necessarily OFDM signals.}
We assume that the channel from PU~2 to SU~1 is a block-fading channel in the time domain. 
That is, all the OFDM symbols in the backward transmissions experience the same channel.
Denote $\mathbf{Y}(l,k)$ as the $l$th sample of the overheard interfering signal on subcarrier $k$ in Phase~I.
Then, we have\footnote{For the notation in this paper, superscripts ``[1]'' and ``[2]'' mean Phases I and II, respectively. Subscripts ``s'' and ``p'' mean the secondary and primary users, respectively.} 
\begin{equation}
\mathbf{Y}(l,k) = \mathbf{H}_\mathrm{sp}^\mathrm{[1]}(k) \mathbf{X}_\mathrm{p}^\mathrm{[1]}(l,k) + \mathbf{W}(l,k),
\label{eq:beamforming_mdoel}
\end{equation}
where $\mathbf{H}_\mathrm{sp}^\mathrm{[1]}(k) \in \mathbb{C}^{M_\mathrm{s} \times M_\mathrm{p}}$ is the matrix representation of the block-fading channel from PU~2 to SU~1 on subcarrier $k$, 
$\mathbf{X}_\mathrm{p}^\mathrm{[1]}(l,k) \in \mathbb{C}^{M_\mathrm{p} \times 1}$ is the interfering signal transmitted by PU~2 on subcarrier $k$, 
and
$\mathbf{W}(l,k) \in \mathbb{C}^{M_\mathrm{s} \times 1}$ is the noise vector at SU~1.
Also, note that SU 1 knows $\mathbf{Y}(l,k)$ but does not know $\mathbf{H}_\mathrm{sp}^\mathrm{[1]}(k)$, $\mathbf{X}_\mathrm{p}^\mathrm{[1]}(l,k)$, and $\mathbf{W}(l,k)$.

At SU~1, we seek a spatial filter that can combine the overheard interfering signals in a destructive manner.
Denote $\mathbf{P}(k)$ as the spatial filter on subcarrier $k$.
Then, the problem of designing $\mathbf{P}(k)$ can be expressed as:
\begin{equation}
\mathrm{min} ~~
\mathbb{E}[\mathbf{P}(k)^* \mathbf{Y}(l,k) \mathbf{Y}(l,k)^* \mathbf{P}(k)], ~~~~ 
\mathrm{s.t.} ~
\mathbf{P}(k)^* \mathbf{P}(k) = 1,
\label{eq:beamforming_opt_objective}
\end{equation}
where $(\cdot)^*$ represents conjugate transpose operator.

\noindent
\textbf{Construction of Spatial Filters.}
To solve the optimization problem in (\ref{eq:beamforming_opt_objective}), we use Lagrange multipliers method.
We define the Lagrange function as:
\begin{equation}
\!
\mathcal{L}(\mathbf{P}(k), \lambda) 
\! = \!
\mathbb{E}\!\big[\!\mathbf{P}(k)^* \mathbf{Y}(l,k) \mathbf{Y}(l,k)^* \mathbf{P}(k)\!\big]
\! - \!
\lambda \big[\mathbf{P}(k)^* \mathbf{P}(k) \!-\! 1\big],
\end{equation}
where $\lambda$ is Lagrange multiplier.

By setting the partial derivatives of $\mathcal{L}(\mathbf{P}(k), \lambda)$ to zero, we have
\begin{align}
&\frac{\partial \mathcal{L}(\mathbf{P}(k), \lambda)}{\partial \mathbf{P}(k)} =  \mathbf{P}(k)^* \Big(\mathbb{E} [\mathbf{Y}(l,k) \mathbf{Y}(l,k)^*] - \lambda \mathbf{I}\Big)  = 0,
\label{eq:lagrange_derivative1}\\
&\frac{\partial \mathcal{L}(\mathbf{P}(k), \lambda)}{\partial \lambda} = \mathbf{P}(k)^* \mathbf{P}(k)  - 1 = 0.
\label{eq:lagrange_derivative2}
\end{align}

Based on the definition of eigendecomposition, it is easy to see that the solutions to equations (\ref{eq:lagrange_derivative1}) and (\ref{eq:lagrange_derivative2}) are the eigenvectors of 
$\mathbb{E}[\mathbf{Y}(l,k) \mathbf{Y}(l,k)^*]$
and the corresponding values of $\lambda$ are the eigenvalues of 
$\mathbb{E}[\mathbf{Y}(l,k) \mathbf{Y}(l,k)^*]$.
Note that $\mathbb{E}[\mathbf{Y}(l,k) \mathbf{Y}(l,k)^*]$ has $M_\mathrm{s}$ eigenvectors, each of which corresponds to a stationary point of the Lagrange function (extrema, local optima, and global optima).  
As $\lambda$ is the penalty multiplier for the Lagrange function, the optimal spatial filter $\mathbf{P}(k)$ lies within the subspace spanned by the eigenvectors of $\mathbb{E}[\mathbf{Y}(l,k) \mathbf{Y}(l,k)^*]$ that correspond to the minimum eigenvalue.

For Hermitian matrix $\mathbb{E}[\mathbf{Y}(l,k) \mathbf{Y}(l,k)^*]$, it may have multiple eigenvectors that correspond to the minimum eigenvalue.
Denote $M_\mathrm{e}$ as the number of eigenvectors that correspond to the minimum eigenvalue. 
Then, we can write them as: 
\begin{equation}
[\mathbf{U}_1, \mathbf{U}_2, \cdots, \mathbf{U}_{M_\mathrm{e}}] = 
\mathit{mineigvectors}\Big(\mathbb{E}[\mathbf{Y}(l,k) \mathbf{Y}(l,k)^*]\Big),
\label{eq:min_eigenvectors}
\end{equation}
where $\mathit{mineigvectors}(\cdot)$ represents the eigenvectors that correspond to the minimum eigenvalue.

To estimate $\mathbb{E}[\mathbf{Y}(l,k) \mathbf{Y}(l,k)^*]$ in (\ref{eq:min_eigenvectors}), we average the received interfering signal samples over the time. 
Denote $\mathbf{Y}(l,k)$ as the $l$th sample of the received interfering signals on subcarrier $k$.
Then, we have
\begin{equation}
[\mathbf{U}_1, \mathbf{U}_2, \cdots, \mathbf{U}_{M_\mathrm{e}}] = 
\mathit{mineigvectors}\Big(\sum_{l=1}^{L_\mathrm{p}}\mathbf{Y}(l,k) \mathbf{Y}(l,k)^*\Big).
\label{eq:calc_eigenvectors}
\end{equation}

Based on (\ref{eq:calc_eigenvectors}), the optimal filter $\mathbf{P}(k)$ can be written as:
\begin{equation}
\mathbf{P}(k) =  \sum_{m = 1}^{M_\mathrm{e}} \alpha_m \mathbf{U}_m,
\label{eq:optimal_precoder}
\end{equation}
where $\alpha_m$ is a weight coefficient with $\sum_{m = 1}^{M_\mathrm{e}} \alpha_m^2 = 1$.

Now, we summarize the BBF technique as follows.
In Phase~I, SU~1 overhears interfering signals $\mathbf{Y}(l,k)$ from PU~2.
Based on the overheard interfering signals, it constructs a spatial filter $\mathbf{P}(k)$ for subcarrier $k$ using
(\ref{eq:calc_eigenvectors}) and (\ref{eq:optimal_precoder}).
In Phase~II, we use $\overline{\mathbf{P}(k)}$ as the precoding vector for beamforming on subcarrier $k$, where $\overline{(\cdot)}$ is the element-wise conjugate operator.
%
%
For this beamforming technique, we have the following remarks:

\textit{Remark 1.}
It is evident that this beamforming technique does not require explicit CSI. 
Instead, it directly uses the received interfering signals to construct the precoding vectors for beamforming. 
Therefore, this technique is termed as ``blind'' beamforming. 

\textit{Remark 2.}
In practice, the noises from SU~1's antennas are typically drawn from independent identical distributions.
If that is the case, the number of eigenvectors in (\ref{eq:calc_eigenvectors}) that correspond to the minimum eigenvalue is $M_\mathrm{e} = M_\mathrm{s} - M_\mathrm{p}$.
Therefore, in (\ref{eq:optimal_precoder}), we have $(M_\mathrm{s} - M_\mathrm{p})$ free variables $\alpha_m$ that can be optimized to maximize the signal strength at the secondary receiver (SU~2).

\textit{Remark 3.}
This beamforming technique only involves  one-time eigendecomposition for each subcarrier. 
It has similar computational complexity as zero-forcing (ZF) and minimum mean square error (MMSE) precoding techniques. 
Therefore, it is amenable to practical implementation.

\noindent
\textbf{IC Capability of BBF.}
For the performance of the proposed beamforming technique, we have the following lemma:

\begin{lemma}
\em
The proposed beamforming technique completely pre-cancels interference at the primary receiver if 
(i)~forward and backward channels are reciprocal;
and
(ii) the noise is zero.
\label{lem:bbf}
\end{lemma}


The proof of Lemma~\ref{lem:bbf} is given in Appendix~\ref{app:bbf}.
To maintain the reciprocity of forward and backward channels in practical wireless systems, we can employ the relative calibration method in \cite{shepard2012argos}. 
This relative calibration method is an internal and standalone calibration method that can be done with assistance from one device (e.g., SU 1).
In our experiment, we implement this calibration method to preserve the channel reciprocity.

\section{Blind Interference Cancellation}
\label{sec:bic}	

In this section, we focus on SU~2 in Phase~II as shown in Fig.~\ref{fig:spectrum_sharing_scheme}(b).
We develop a BIC technique for the secondary receiver (SU~2) to decode its desired signals in the presence of interference from the primary transmitter (PU~1).

\noindent
\textbf{Mathematical Formulation.}
Recall that we use IEEE~802.11 legacy PHY for data transmission in the secondary network.
Specifically, SU~1 sends packet-based signals to SU~2, which comprise a bulk of OFDM symbols.
In each packet, the first four OFDM symbols carry preambles (pre-defined reference signals) and the remaining OFDM symbols carry payloads.

Consider the signal transmission in Fig.~\ref{fig:spectrum_sharing_scheme}(b).
Denote $X_\mathrm{s}^\mathrm{[2]}(l, k)$ as the signal that SU~1 transmits on subcarrier~$k$ in OFDM symbol $l$.
Denote $\mathbf{X}_\mathrm{p}^\mathrm{[2]}(l, k)$ as the signal that PU~1 transmits on subcarrier $k$ in OFDM symbol~$l$.\footnote{PU~1 does not necessarily send OFDM signals. But at SU~2, the interfering signals from PU~1 can always be converted to the frequency domain using FFT operation.}
Denote $\mathbf{Y}(l,k)$ as the received signal vector at SU~2 on subcarrier $k$ in OFDM symbol~$l$.
Then, we have 
\begin{equation}
	\mathbf{Y}(l, k) = 
	\mathbf{H}_\mathrm{ss}^\mathrm{[2]}(k) \overline{\mathbf{P}(k)} X_\mathrm{s}^\mathrm{[2]}(l, k) +
	\mathbf{H}_\mathrm{sp}^\mathrm{[2]}(k) \mathbf{X}_\mathrm{p}^\mathrm{[2]}(l, k) + \mathbf{W}(l, k), 
	\label{eq:mimo_system_model}
\end{equation}
where 
$\mathbf{H}_\mathrm{ss}^\mathrm{[2]}(k)$ is the block-fading channel between SU~2 and SU~1 on subcarrier~$k$,
$\mathbf{H}_\mathrm{sp}^\mathrm{[2]}(k)$ is the block-fading channel between SU~2 and PU~1 on subcarrier~$k$,
and
$\mathbf{W}(l,k)$ is the noise on subcarrier~$k$ in OFDM symbol~$l$.

At SU 2, in order to decode the intended signal in the presence of cross-network interference, we use a linear spatial filter $\mathbf{G}(k)$ for all OFDM symbols on subcarrier~$k$. 
Then, the decoded signal can be written as:
\begin{equation}
	\hat{X}_\mathrm{s}^\mathrm{[2]}(l, k) = \mathbf{G}(k)^* \mathbf{Y}(l, k).
	\label{eq:decoded_signal}
\end{equation}

While there exist many criteria for the design of $\mathbf{G}(k)$, our objective is to minimize the mean square error (MSE) between the decoded and original signals.
Thus, the signal detection problem can be formulated as:
\begin{equation}
	\min~~ \mathbb{E} \Big[ \left|\hat{X}_\mathrm{s}^\mathrm{[2]}(l, k) - X_\mathrm{s}^\mathrm{[2]}(l, k) \right|^2 \Big].
	\label{eq:opt_bic}
\end{equation}

\noindent
\textbf{Construction of Spatial Filters.}
To solve the optimization problem in (\ref{eq:opt_bic}), we use Lagrange multipliers method again.
We define the Lagrange function as:
\begin{equation}
	\mathcal{L}(\mathbf{G}(k)) 
	=
	\mathbb{E} \Big[ \left|\hat{X}_\mathrm{s}^\mathrm{[2]}(l, k) - X_\mathrm{s}^\mathrm{[2]}(l, k) \right|^2 \Big].
	\label{eq:bic_lagrange}
\end{equation}

Based on (\ref{eq:decoded_signal}), (\ref{eq:bic_lagrange}) can be rewritten as:
\begin{equation}
	\mathcal{L}(\mathbf{G}(k)) 
	=
	\mathbb{E} \Big[ \left|\mathbf{G}(k)^* \mathbf{Y}(l, k) - X_\mathrm{s}^\mathrm{[2]}(l, k) \right|^2 \Big].
	\label{eq:bic_lagrange2}
\end{equation}

Equation (\ref{eq:bic_lagrange2}) is a quadratic function of $\mathbf{G}(k)$. 
To minimize MSE, we can take the gradient with respect to $\mathbf{G}(k)$.
The optimal filter $\mathbf{G}(k)$ can be obtained by setting the gradient to zero, which we show as follows:
\begin{equation}
	\mathbb{E} \big[\mathbf{Y}(l,k)\mathbf{Y}(l,k)^*\big] \mathbf{G	}(k) - \mathbb{E} \big[\mathbf{Y}(l,k) X_\mathrm{s}^\mathrm{[2]}(l, k)^*\big]
	= 
	0.
	\label{eq:lagrange_derivative_bic}
\end{equation}

\begin{figure}
	\centering
	\includegraphics[width=3.4in]{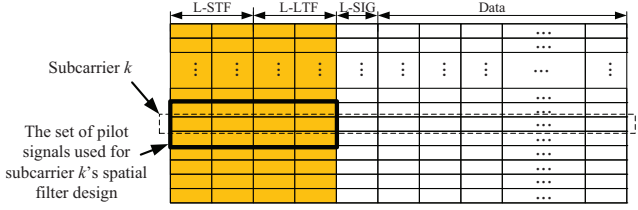}
	\caption{An example of $\mathcal{Q}(k)$ in IEEE~802.11 legacy frame.}
	\label{fig:frame}
\end{figure}

Based on (\ref{eq:lagrange_derivative_bic}), we obtain the optimal filter
\begin{equation}
	\mathbf{G}(k) = \mathbb{E} \big[\mathbf{Y}(l,k)\mathbf{Y}(l,k)^*\big]^{+} \mathbb{E} \big[\mathbf{Y}(l,k) X_\mathrm{s}^\mathrm{[2]}(l, k)^*\big] \;,
	\label{eq:optimalG}
\end{equation}
where  $(\cdot)^{+}$ denotes pseudo inverse operation. Equation~(\ref{eq:optimalG}) is the optimal design of $\mathbf{G}(k)$ in the sense of minimizing MSE. 
To calculate $\mathbb{E} \big[\mathbf{Y}(l,k)\mathbf{Y}(l,k)^*\big]$ and $\mathbb{E} \big[\mathbf{Y}(l,k) X_\mathrm{p}^\mathrm{[2]}(l,k)^*\big]$ in~(\ref{eq:optimalG}), we can take advantage of the pilot (reference) symbols in wireless systems  (e.g., the preamble in IEEE~802.11 legacy frame).
Denote $\mathcal{Q}_k$ as the set of pilot symbols in a frame that can be used for the design of interference mitigation filter $\mathbf{G}(k)$. 
Then, we can approach the statistical expectations in (\ref{eq:optimalG}) using the averaging operations as follows:
\begin{equation}
	\mathbb{E} \big[\mathbf{Y}(l,k)\mathbf{Y}(l,k)^*\big] 
	\approx
	\frac{1}{|\mathcal{Q}_k|}
	\!\! 
	\sum_{(l,k') \in \mathcal{Q}_k}
	\!\!\!\!\! 
	{\mathbf{Y}}(l,k'){\mathbf{Y}}(l,k')^* \;,
	\label{eq:YY}
\end{equation}
\begin{equation}
	\mathbb{E} \big[\mathbf{Y}(l,k) X_\mathrm{p}^\mathrm{[2]}(l,k)^*\big] 
	\approx
	\frac{1}{|\mathcal{Q}_k|}
	\!\!\!\!  
	\sum_{(l,k') \in \mathcal{Q}_k}
	\!\!\!\!\!\!\! 
	{\mathbf{Y}}(l,k'){ X_\mathrm{p}^\mathrm{[2]}}(l,k')^*  ,
	\label{eq:YX}
\end{equation}
where an example of $\mathcal{Q}_k$ is illustrated in Fig.~\ref{fig:frame}.

Note that, with a bit abuse of notation, we replaced the approximation sign in (\ref{eq:YY}) and (\ref{eq:YX}) with an equation sign for simplicity.
Then, the spatial filter $\mathbf{G}(k)$ can be written as:
\begin{align}
	\mathbf{G}(k)
	\!=\!\!
	\Big[
	\!\!\!\!\! 
	\sum_{(l,k') \in \mathcal{Q}_k} 
	\!\!\!\!\! \!
	{\mathbf{Y}}(l,k') {\mathbf{Y}}(l,k')^*\Big]^{+}
	\Big[
	\!\!\!\!\! 
	\sum_{(l,k') \in \mathcal{Q}_k} 
	\!\!\!\!\! 
	{\mathbf{Y}}(l,k') {X_\mathrm{p}^\mathrm{[2]}(l,k')}^*\Big].
	\label{eq:bic_filter}
\end{align}

We now summarize our BIC technique as follows.
In Phase~II, SU~2 needs to decode its desired signal in the presence of interference from PU~1. 
To do so, SU~2 first constructs a spatial filter for each of its subcarriers using (\ref{eq:bic_filter}), and then decodes its desired signal using (\ref{eq:decoded_signal}). 
For this BIC technique, we have the following remarks:

\textit{Remark 4.}
The spatial filter in (\ref{eq:bic_filter}) not only cancels the interference but also equalizes the channel distortion for signal detection. 

\textit{Remark 5.}
As shown in (\ref{eq:bic_filter}) and (\ref{eq:decoded_signal}), our BIC technique does not require knowledge about the interference characteristics, including waveform and bandwidth.
That's the reason that it is referred to as ``blind'' interference cancellation.

\textit{Remark 6.}
This BIC scheme does not require explicit CSI.
Rather, it only requires pilot signals at the secondary transmitter. 
In contrast to conventional signal detection schemes, such as ZF and MMSE detectors, the BIC technique technique does not require channel estimation. 

\textit{Remark 7.}
As shown in (\ref{eq:bic_filter}) and (\ref{eq:decoded_signal}), this BIC technique involves matrix inversion and multiplication, where the dimension of the matrix equals to the number of antennas on a secondary user. 
Its computational complexity is similar to that of ZF detection technique, which is widely used in real-world wireless systems.
Therefore, we do not expect computational issue with this BIC technique.

\begin{figure}
	\centering
	\begin{subfigure}[b]{0.15\textwidth}
		\centering
		\includegraphics[width=1.15in]{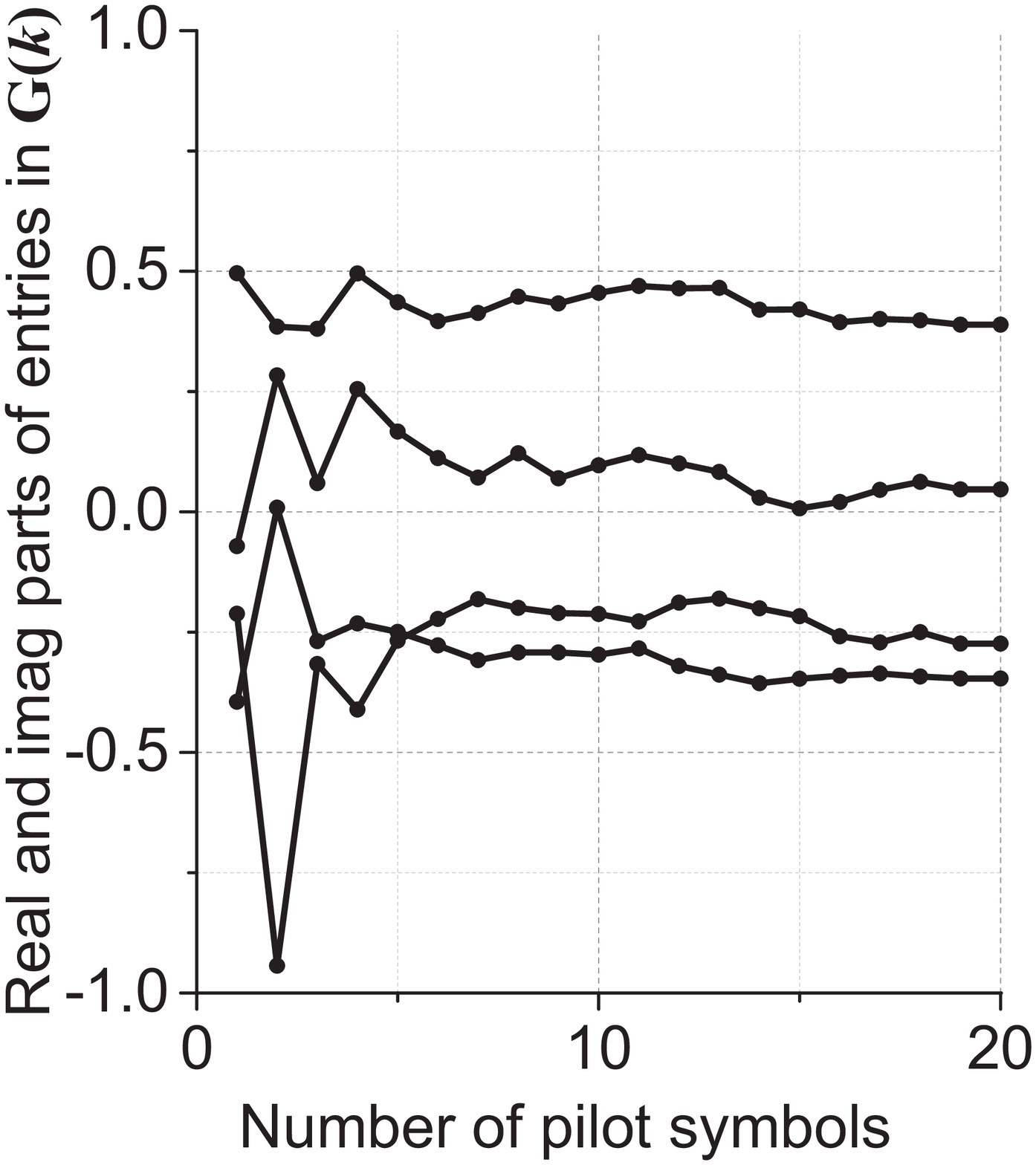}
		\caption{SNR=5dB case}
	\end{subfigure}
	~
	\begin{subfigure}[b]{0.15\textwidth}
		\centering
		\includegraphics[width=1.15in]{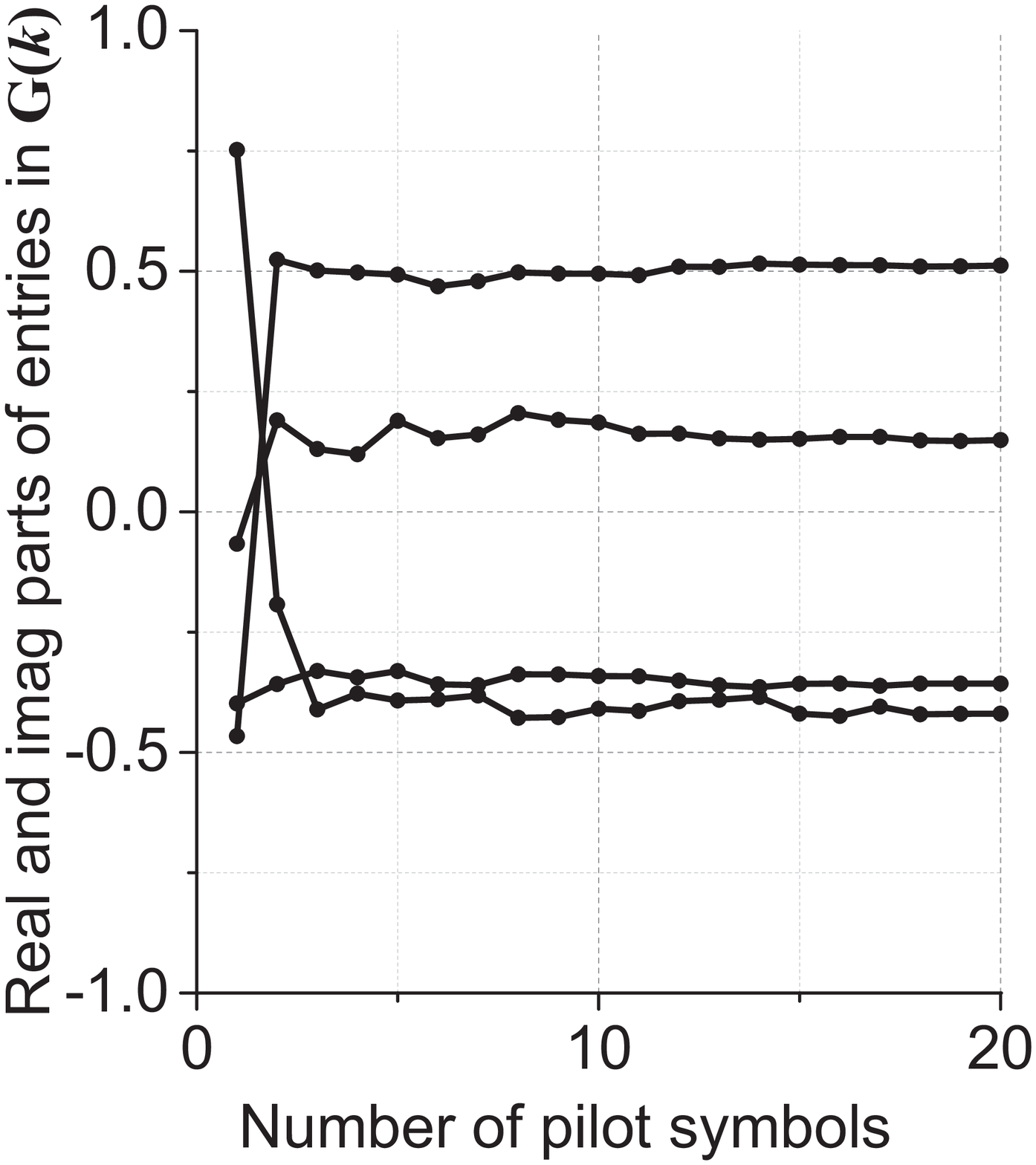}
		\caption{SNR=15dB case}
	\end{subfigure}
	~
	\begin{subfigure}[b]{0.15\textwidth}
		\centering
		\includegraphics[width=1.15in]{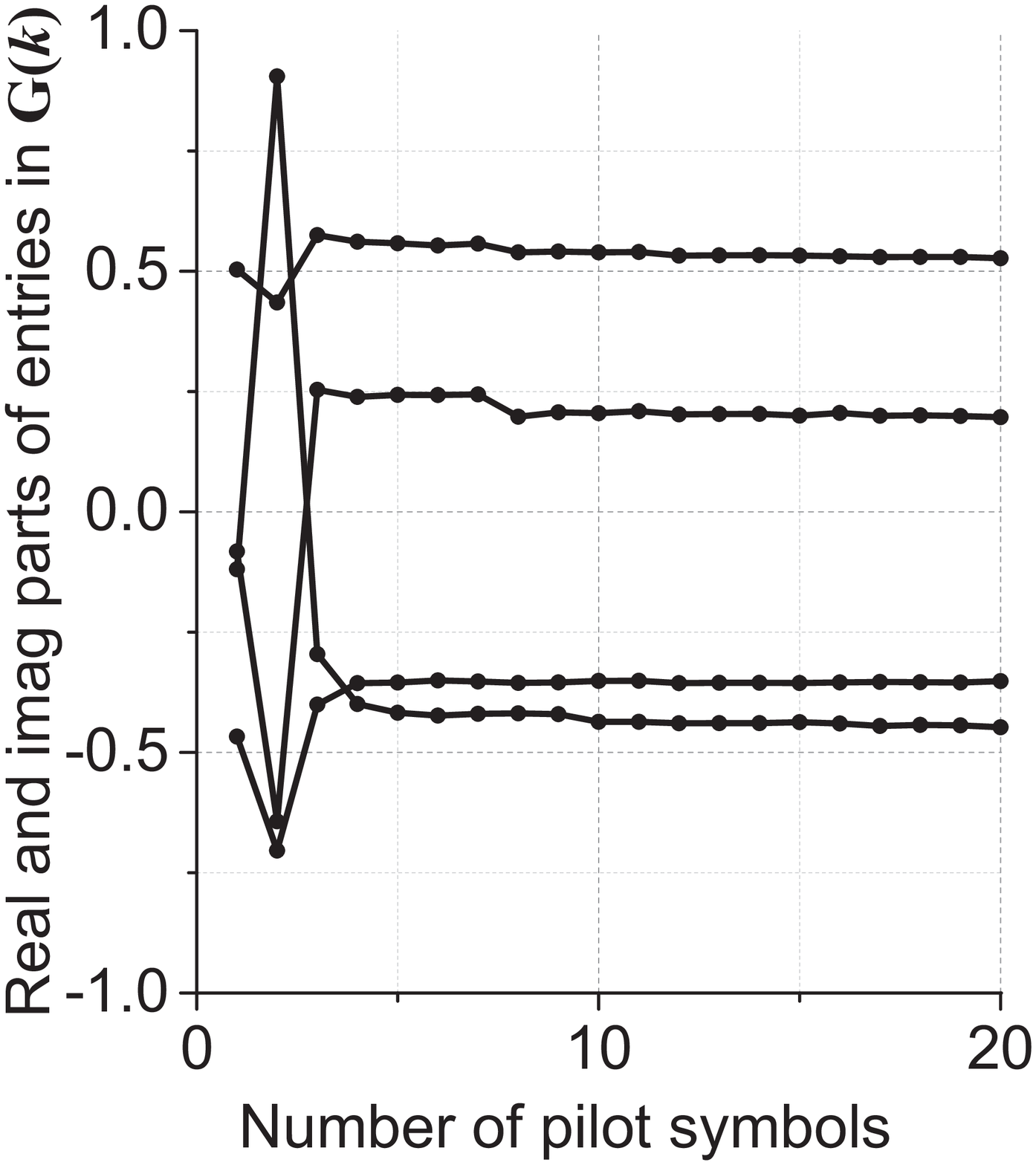}
		\caption{SNR=25dB case}
	\end{subfigure}		
	\caption{Convergence speed of spatial filter over the number of pilot symbols in $(M_\mathrm{p}=1, M_\mathrm{s} = 2)$ network.}
	\label{fig:convergence_2ue}
\end{figure}
\begin{figure}
	\centering
	\begin{subfigure}[b]{0.15\textwidth}
		\centering
		\includegraphics[width=1.15in]{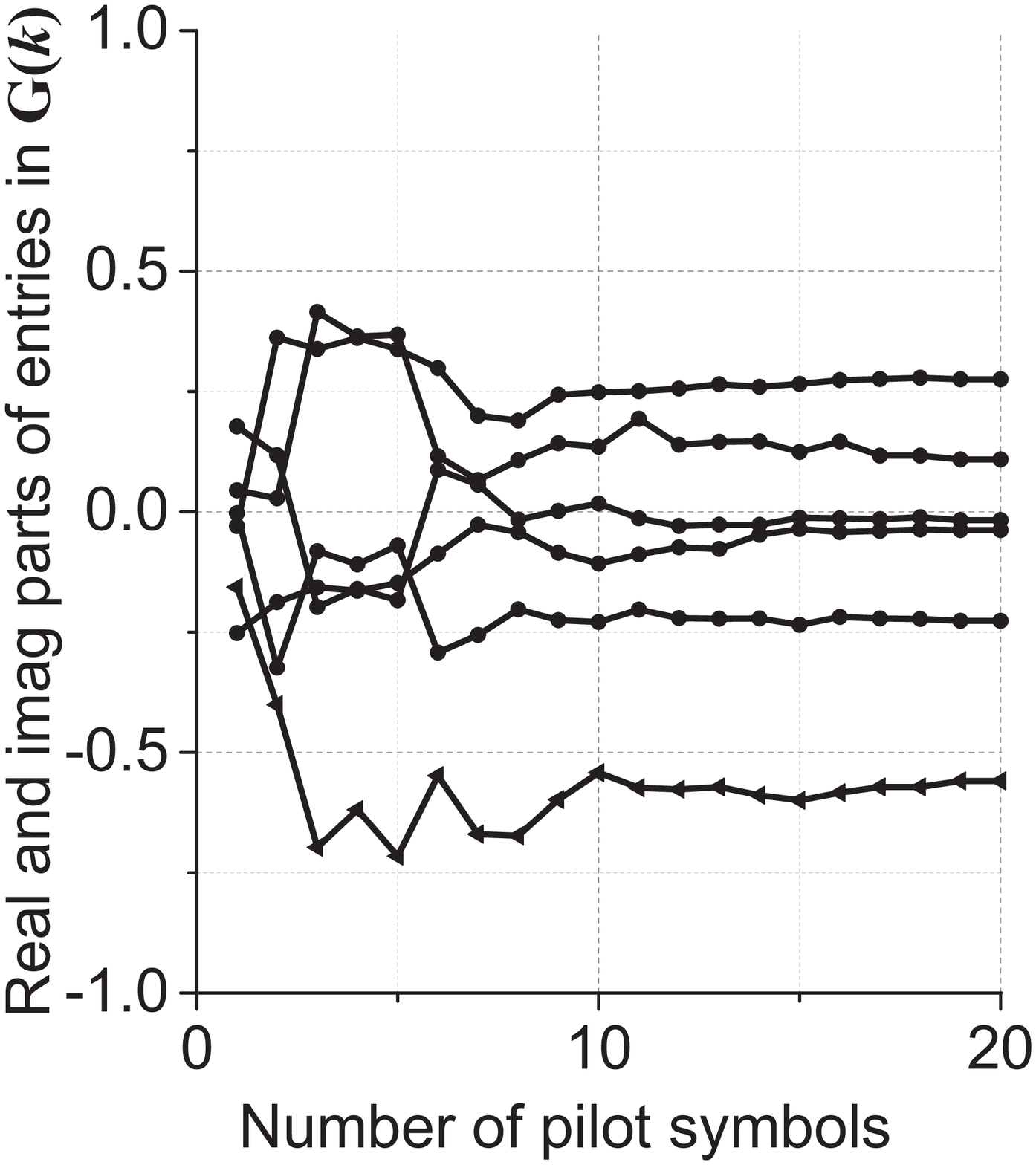}
		\caption{SNR=5dB case}
	\end{subfigure}
	~
	\begin{subfigure}[b]{0.15\textwidth}
		\centering
		\includegraphics[width=1.15in]{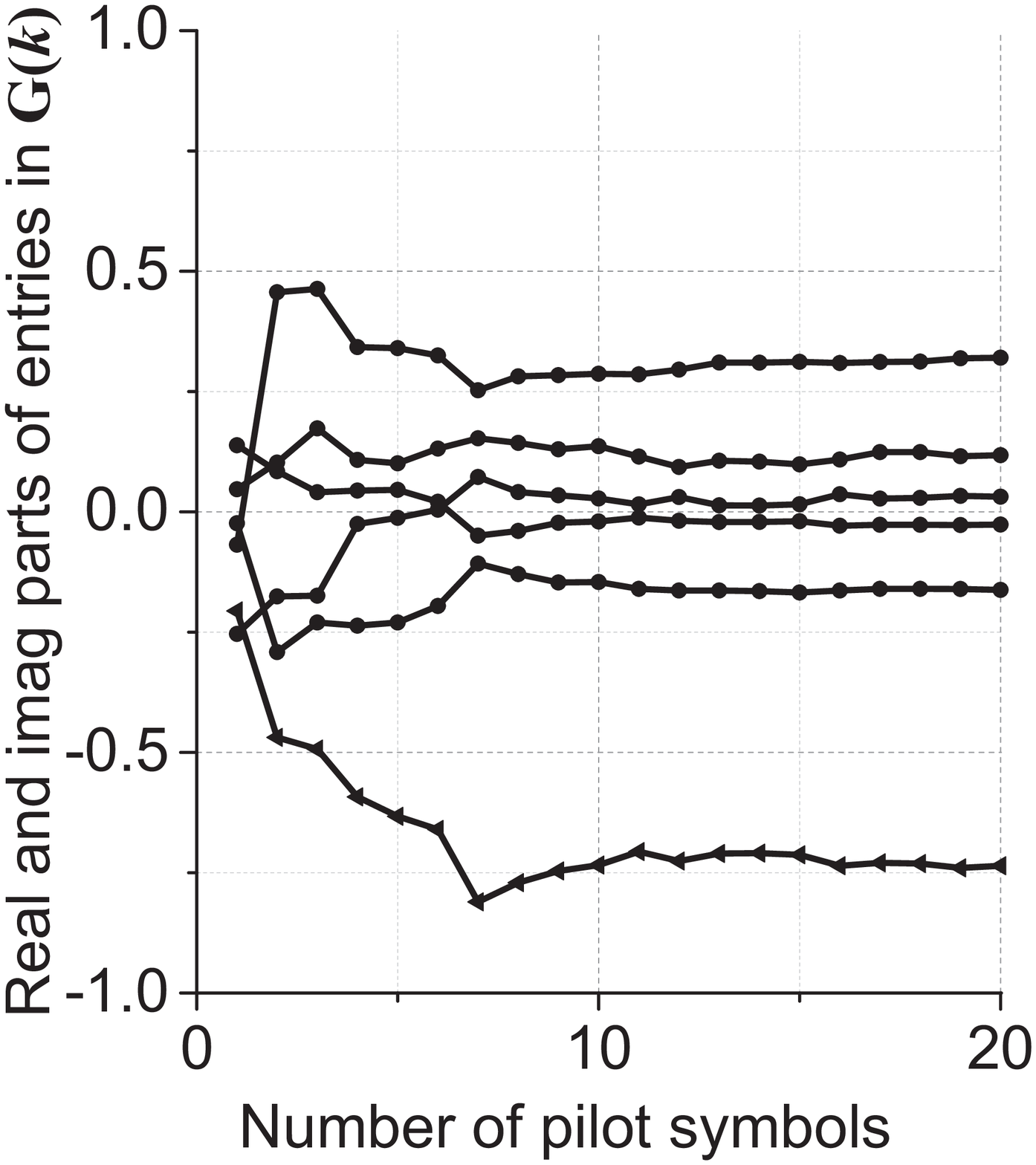}
		\caption{SNR=15dB case}
	\end{subfigure}
	~
	\begin{subfigure}[b]{0.15\textwidth}
		\centering
		\includegraphics[width=1.15in]{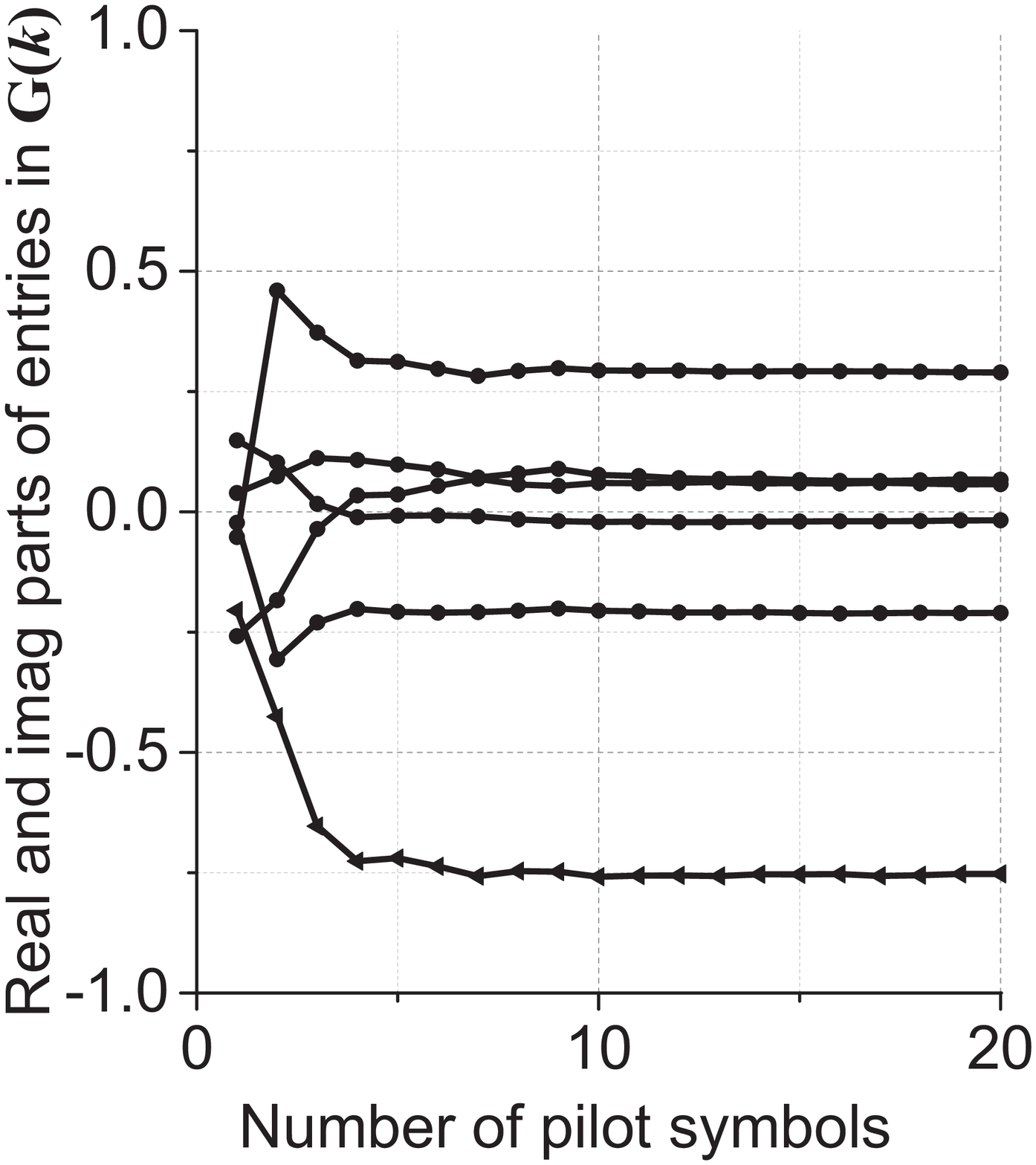}
		\caption{SNR=25dB case}
	\end{subfigure}		
	\caption{Convergence speed of spatial filter over the number of pilot symbols in $(M_\mathrm{p}=2, M_\mathrm{s} = 3)$ network.}
	\label{fig:convergence_3ue}
\end{figure}

\noindent
\textbf{IC Capability of BIC.}
For the performance of the proposed spatial method, we have the following lemma:
\begin{lemma}
	\em
	If the pilot signals are sufficient and the noise is zero, the BIC scheme can perfectly recover the signals in the presence of cross-network interference (i.e., $\hat X_\mathrm{s}^\mathrm{[2]}(k, l) = X_\mathrm{s}^\mathrm{[2]}(k, l)$, $\forall k, l$) .
	\label{lem:bic}
\end{lemma} 

The proof of Lemma~\ref{lem:bic} is given in Appendix~\ref{app:bic}.

\noindent
\textbf{Pilot Signals for Spatial Filter Construction.}
Lemma~\ref{lem:bic} shows the superior performance of our BIC technique when the pilot signals are sufficient. 
A natural question is how many pilot signals on each subcarrier is considered sufficient.
To answer this question, we first present our simulation results to study the convergence speed of the spatial filter over the number of pilot signals, and then propose a method to increase the number of pilot signals for the spatial filter construction.

As an instance, we simulated the convergence speed of the spatial filter over the number of pilot symbols for SU~2 in Fig.~\ref{fig:spectrum_sharing_scheme}.
Fig.~\ref{fig:convergence_2ue} and Fig.~\ref{fig:convergence_3ue} respectively present our simulation results in two network settings: $(M_\mathrm{p}=1, M_\mathrm{s} = 2)$ and $(M_\mathrm{p}=2, M_\mathrm{s} = 3)$.
From the simulation results, we can see that the spatial filter converges at a pretty fast speed in these two network settings. 
Specifically, the spatial filter can achieve a good convergence within about $10$ pilot signal symbols. 


Recall that the secondary network uses IEEE~802.11 legacy frame for transmission from SU~1 to SU~2, which only has four pilot symbols on each subcarrier (i.e., two L-STF OFDM symbols and two L-LTF OFDM symbols). 
So, the construction of spatial filter is in shortage of pilot symbols. 
To address this issue, for each subcarrier, we not only use the pilot symbols on that subcarrier but also the pilot symbols on its neighboring subcarriers, as illustrated in Fig.~\ref{fig:frame}.
The rationale behind this operation lies in the fact that channel coefficients on neighboring subcarriers are highly correlated in real-world wireless environments.
By leveraging the pilot symbols on two neighboring subcarriers, we have $12$ pilot symbols for the construction of the spatial filter, which appears to be sufficient based on our simulation results in Fig.~\ref{fig:convergence_2ue} and Fig.~\ref{fig:convergence_3ue}.
We note that analytically studying the performance of BIC with respect to the number and format of pilot signals is beyond the scope of this work. 
Instead, we resort to experiments to study its performance in real network settings.



\section{Performance Evaluation}
\label{sec:evaluation}	
In this section, we consider an underlay CRN in two time slots as shown in Fig.~\ref{fig:experiemental_cases}.
We have built a prototype of the proposed underlay spectrum sharing scheme in this network on a software-defined radio (SDR) testbed and evaluated its performance in real-world indoor wireless environments.

\subsection{Implementation}


\begin{figure}[t]
	\centering
	\begin{subfigure}[b]{1.62in}
		\includegraphics[width=1.70in]{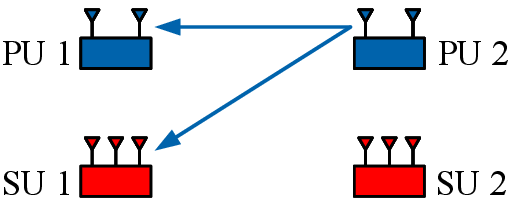}
		\caption{Transmission in phase~I.}
	\end{subfigure}
	~~
	\begin{subfigure}[b]{1.62in}
		\includegraphics[width=1.70in]{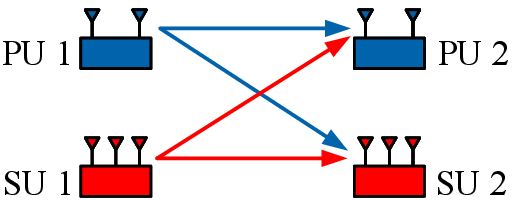}
		\caption{Transmission in Phase~II.}
	\end{subfigure}
	\caption{Experimental setup for an underlay CRN with two network settings: $(M_p \!\!=\!\! 1, M_s \!\!=\!\! 2)$ and $(M_p \!\!=\!\! 2, M_s \!\!=\!\! 3)$.} 
	\label{fig:experiemental_cases}
\end{figure}


\begin{table}[]
	\centering
	\footnotesize
	\caption{The implementation parameters of primary and secondary networks.}
	\begin{tabular}{ccccc}
		\hline
		\multicolumn{1}{c}{} & \multicolumn{1}{c}{\begin{tabular}[c]{@{}c@{}}Primary\\ network 1\end{tabular}} & \multicolumn{1}{c}{\begin{tabular}[c]{@{}c@{}}Primary\\ network 2\end{tabular}} & \multicolumn{1}{c}{\begin{tabular}[c]{@{}c@{}}Primary\\ network 3\end{tabular}} & \multicolumn{1}{c}{\begin{tabular}[c]{@{}c@{}}Secondary\\ network\end{tabular}} \\ \hline
		\multicolumn{1}{c}{\begin{tabular}[c]{@{}c@{}} System\\ type\end{tabular}}   & \multicolumn{1}{c}{Commercial}  & \multicolumn{1}{c}{\begin{tabular}[c]{@{}c@{}} Custom-\\ built\end{tabular}}  & \multicolumn{1}{c}{\begin{tabular}[c]{@{}c@{}} Custom-\\ built\end{tabular}}  & \multicolumn{1}{c}{\begin{tabular}[c]{@{}c@{}} Custom-\\ built\end{tabular}}  \\ \hline
		\multicolumn{1}{c}{Standard} & \multicolumn{1}{c}{Wi-Fi}   & \multicolumn{1}{c}{LTE-like}& \multicolumn{1}{c}{CDMA-like}   & \multicolumn{1}{c}{Wi-Fi-like}  \\ \hline
		\multicolumn{1}{c}{Waveform} & \multicolumn{1}{c}{OFDM}& \multicolumn{1}{c}{OFDM}& \multicolumn{1}{c}{CDMA}& \multicolumn{1}{c}{OFDM}\\ \hline
		\multicolumn{1}{c}{FFT-Point}& \multicolumn{1}{c}{64}  & \multicolumn{1}{c}{1024}& \multicolumn{1}{c}{-}   & \multicolumn{1}{c}{64}  \\ \hline
		\multicolumn{1}{c}{\begin{tabular}[c]{@{}c@{}}Valid\\ subcarriers\end{tabular}}  & \multicolumn{1}{c}{52}  & \multicolumn{1}{c}{600} & \multicolumn{1}{c}{-}   & \multicolumn{1}{c}{52}  \\ \hline
		\multicolumn{1}{c}{\begin{tabular}[c]{@{}c@{}}Sample\\ rate\end{tabular}} & \multicolumn{1}{c}{20 Msps} & \multicolumn{1}{c}{10 Msps} & \multicolumn{1}{c}{5 Msps}  & \multicolumn{1}{c}{\begin{tabular}[c]{@{}c@{}} 5~Mbps,\\ 25~Mbps\end{tabular}}  \\ \hline
		\multicolumn{1}{c}{\begin{tabular}[c]{@{}c@{}}Signal\\ bandwidth\end{tabular}} & \multicolumn{1}{c}{$\sim$16 MHz} & \multicolumn{1}{c}{$\sim$5.8 MHz} & \multicolumn{1}{c}{$\sim$5 MHz}  & \multicolumn{1}{c}{\begin{tabular}[c]{@{}c@{}} $\sim$4.06 MHz\\ $\sim$20.31 MHz\end{tabular}}  \\ \hline
		\multicolumn{1}{c}{\begin{tabular}[c]{@{}c@{}}Carrier \\ frequency\end{tabular}} & \multicolumn{1}{c}{2.48 GHz}& \multicolumn{1}{c}{2.48 GHz}& \multicolumn{1}{c}{2.48 GHz}& \multicolumn{1}{c}{2.48 GHz}\\ \hline
		\multicolumn{1}{c}{\begin{tabular}[c]{@{}c@{}}Max tx \\ power\end{tabular}} & \multicolumn{1}{c}{$\sim$20 dBm}& \multicolumn{1}{c}{$\sim$15 dBm}& \multicolumn{1}{c}{$\sim$15 dBm}& \multicolumn{1}{c}{$\sim$15 dBm}\\ \hline
		\multicolumn{1}{c}{\begin{tabular}[c]{@{}c@{}}Antenna\\ number\end{tabular}} & \multicolumn{1}{c}{1}   & \multicolumn{1}{c}{1, 2}& \multicolumn{1}{c}{1}& \multicolumn{1}{c}{2, 3}\\ \hline
	\end{tabular}
	\label{tab:parameters}
\end{table}

\noindent
\textbf{PHY Implementation.}
We consider three different primary networks:
a COTS Wi-Fi network, a LTE-like network, and a CDMA-like network. 
The commercial Wi-Fi devices are Alfa AWUS036NHA Wireless B/G/N USB Adaptors (802.11n), which have one antenna for radio signal transmission and reception.
The LTE-like and CDMA-like primary devices as well as the secondary devices are built using USRP N210 devices and general-purpose computers.
The USRP devices are used for radio signal transmission/reception while the computers are used for baseband signal processing and MAC protocol implementation (in C++ language).
The implementation parameters are listed in Table~\ref{tab:parameters}.

\noindent
\textbf{MAC Implementation.}
We implement the MAC protocol in Fig.~\ref{fig:mac_protocol} for the primary and secondary networks. 
The packet transmissions in the two networks are aligned in the time domain, as shown in Fig.~\ref{fig:mac_protocol}. 
Since bi-directional communication in the secondary network is symmetric, we only consider the forward communication (from SU~1 to SU~2) in our experiments. 
We implement BBF on SU~1 to avoid the interference from the secondary transmitter to the primary receiver.
We also implement BIC on SU~2 to decode its desired signal from SU~1 in the presence of interference from PU~1.
In addition, we implement the RF chain calibration method~\cite{shepard2012argos} on the secondary user (SU~1 in Fig.~\ref{fig:experiemental_cases}) to maintain the relative channel reciprocity. 
Note that the calibration needs to be done at a low frequency (0.1~Hz in our experiments) and therefore would not consume much airtime resources.

\begin{figure}[t]
	\centering
	\includegraphics[width=3.5in]{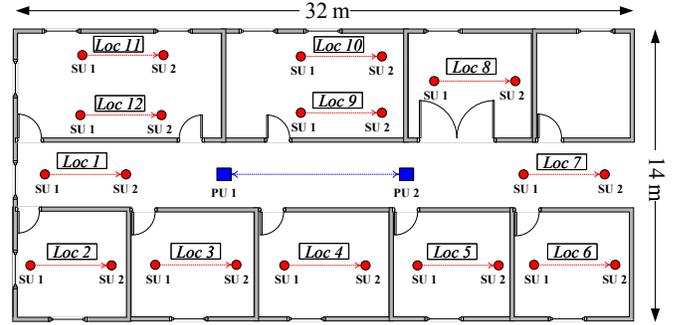}
	\caption{Floor plan of primary and secondary users' locations.} 
	\label{fig:floor_plan}
\end{figure}

\subsection{Experimental Setup and Performance Metrics}

\noindent
\textbf{Experimental Setup.}
Consider the primary and secondary networks in Fig.~\ref{fig:experiemental_cases}.
We place the devices on a floor plan as shown in Fig.~\ref{fig:floor_plan}.
The two primary users are always placed at the spots marked ``PU 1'' and ``PU 2''.
The two secondary users are placed at one of the 12 different locations.
The distance between PU 1 and PU 2 is 10 m and the distance between SU~1 and SU 2 is 6~m. 
The transmit power of primary users is fixed to the maximum level specified in Table~\ref{tab:parameters}, while the transmit power of secondary users is properly adjusted to ensure that its generated interference to the primary receiver (after BBF) is at the noise level.

\noindent
\textbf{Performance Metrics.}
We evaluate the performance of the proposed spectrum sharing scheme using the following four metrics:
\begin{itemize}
	\item Tx-side IC capability at SU~1:  
	This IC capability is from SU~1's BBF.
	It is defined as $\beta_\mathrm{tx} = 10 \log_{10}(P_1/P_2)$, where $P_1$ is the received interference power at PU~2 when SU~1 uses $[\frac{1}{\sqrt{2}}~\frac{1}{\sqrt{2}}]$ or $[\frac{1}{\sqrt{3}}~\frac{1}{\sqrt{3}}~\frac{1}{\sqrt{3}}]$ as the precoder, 
	and $P_2$ is the received interference power at PU~2 when SU~1 uses the precoder constructed by our proposed BBF.
	\item Rx-side IC capability at SU~2:
	This IC capability is from SU~2's BIC. 
	It is defined as $\beta_\mathrm{rx} =\left|\mathrm{EVM}\right| - \max\{\mathit{SIR}_m\}$, where $\mathit{SIR}_m$ is the signal to interference ratio (SIR) on SU~2's $m$th antenna and EVM will be defined in the following. 
	\item Error vector magnitude (EVM) of the decoded signals at SU~2:
	It is defined as follows:
	\begin{align}
	\mathrm{EVM} = 10 \log_{10} \left(\frac{\mathbb{E} \big[\big|\hat{X}_\mathrm{s}^\mathrm{[2]}(l, k) - X_\mathrm{s}^\mathrm{[2]}(l, k)\big|^2\big]}{\mathbb{E} \big[\big|X_\mathrm{s}^\mathrm{[2]}(l, k)\big|^2\big]}
	\right).
	\label{eq:evm_definition}
	\end{align}
	\item Throughput of the secondary network:
	The throughput is extrapolated based on the measured EVM at SU~2 and the modulation and coding scheme (MCS) specified in IEEE~802.11ac standard \cite{IEEE80211ac}.
	Specifically, it is calculated as follows:
	\begin{equation}
	r = \frac{1}{2} \cdot \frac{48}{80} \cdot b \cdot \gamma(\mathrm{EVM}),
	\label{eq:evm_tp}
	\end{equation}
	where $1/2$ is the half-time use of spectrum, 48 is the number of valid subcarriers, 80 is the number of samples in an OFDM symbol, $b$ is the bandwidth, and $\gamma(\mathrm{EVM})$ is the average number of bits carried by one subcarrier and it is given in Table~\ref{tab:evm}. 
\end{itemize}
\begin{table}
	\setstretch{1.25}
	\tiny
	\centering
	\caption{EVM specification in IEEE 802.11ac standard \cite{IEEE80211ac}.}
	\label{tab:evm}
	\begin{tabular}{|l|c|c|c|c|c|c|c|c|c|c|c|}
		\hline
		\!\!\!\!\!\!\!\!\! $\mbox{EVM}$ (dB)\!\!\!\!\!\!\!\!\!    &    \!\!\!\!\!\!(inf~-5)\!\!\!\!\!\!       & \!\!\!\!\!\![-5~-10)\!\!\!\!\!\!   & \!\!\!\!\!\![-10~-13)\!\!\!\!\!\!  & \!\!\!\!\!\![-13~-16)\!\!\!\!\!\!  & \!\!\!\!\!\![-16~-19)\!\!\!\!\!\!   & \!\!\!\!\!\![-19~-22)\!\!\!\!\!\!   & \!\!\!\!\!\![-22~-25)\!\!\!\!\!\!   & \!\!\!\!\!\![-25~-27)\!\!\!\!\!\!   & \!\!\!\!\!\![-27~-30)\!\!\!\!\!\!   & \!\!\!\!\!\![-30~-32)\!\!\!\!\!\!    & \!\!\!\!\!\![-32~-inf)\!\!\!\!\!\!    \\ \hline
		\!\!\!\!\!\!{Modulation}\!\!\!\!\!\!                 &\!\!\!\!\!\!\! N/A \!\!\!\!\!\!\!&\!\!\!\!\!\!\! BPSK \!\!\!\!\!\!\!&\!\!\!\!\!\!\! QPSK \!\!\!\!\!\!\!&\!\!\!\!\!\!\! QPSK \!\!\!\!\!\!\!&\!\!\!\!\!\!\! 16QAM \!\!\!\!\!\!\!&\!\!\!\!\!\!\! 16QAM \!\!\!\!\!\!\!&\!\!\!\!\!\!\! 64QAM \!\!\!\!\!\!\!&\!\!\!\!\!\!\! 64QAM \!\!\!\!\!\!\!&\!\!\!\!\!\!\! 64QAM \!\!\!\!\!\!\!&\!\!\!\!\!\!\! 256QAM \!\!\!\!\!\!\!&\!\!\!\!\!\!\! 256QAM\!\!\!\!\!\! \\ \hline
		\!\!\!\!\!\!{Coding rate}\!\!\!\!\!\!             & \!\!\!\!\!\!\! N/A \!\!\!\!\!\!\!& 1/2  & 1/2  & 3/4  & 1/2   & 3/4   & 2/3   & 3/4   & 5/6   & 3/4    & 5/6    \\ \hline
		\!\!\!\!\!\!{$\gamma$($\mbox{EVM}$)}\!\!\!\!\!\!  & 0 & 0.5  & 1    & 1.5  & 2     & 3     & 4     & 4.5   & 5     & 6      & 20/3   \\ \hline
	\end{tabular}
\end{table}
 
 \begin{figure} 
 	\centering
 	\begin{subfigure}[t]{0.23\textwidth}
 		\centering
 		\includegraphics[width=40mm]{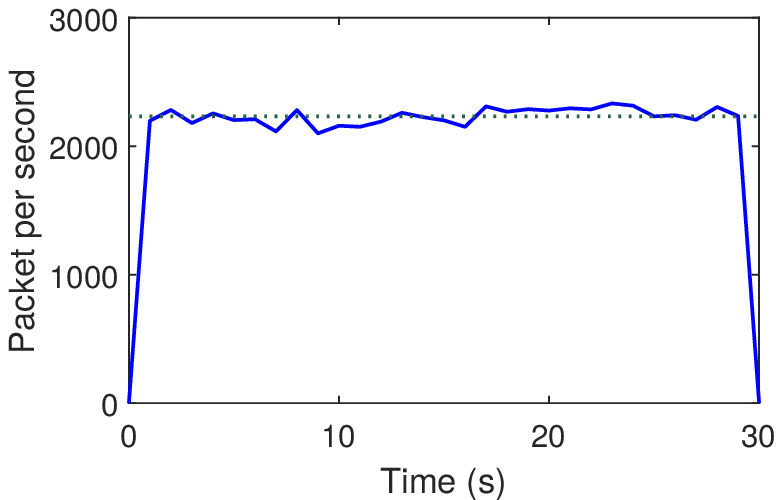}
 		\caption{Interference-free scenario.}
 	\end{subfigure}	
 	~~
 	\begin{subfigure}[t]{0.23\textwidth}
 		\centering
 		\includegraphics[width=40mm]{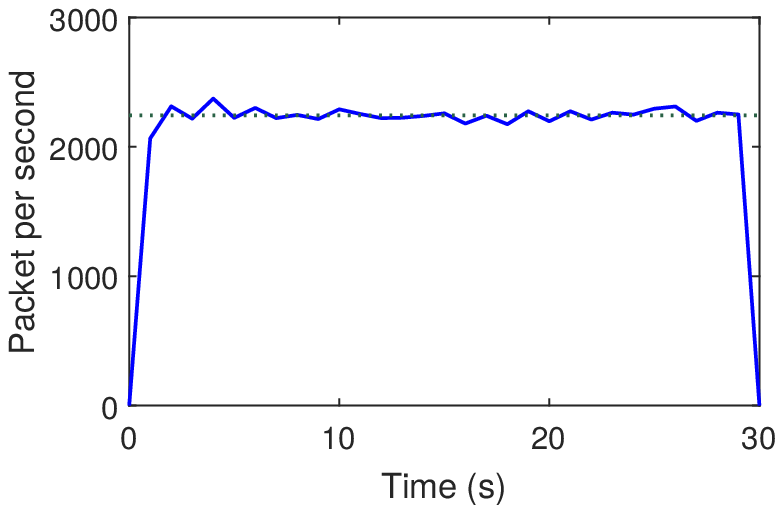}
 		\caption{Spectrum sharing scenario.}
 	\end{subfigure}		
 	\caption{Packet delivery rate in the primary network.} 
 	\label{fig:comerc_primary}
 \end{figure}

\subsection{Coexistence with Commercial Wi-Fi Devices}

We consider primary network 1 in Table~\ref{tab:parameters}.
The two primary devices are set up using commercial Wi-Fi adapters installed on computers, each of which is equipped with one antenna. 
The two primary devices are connected in the ad-hoc mode, and they send data packets to each other as shown in Fig.~\ref{fig:mac_protocol}.
These two primary devices are placed at the spot marked by blue squares in Fig.~\ref{fig:floor_plan}.
The secondary network used in this case is also specified in Table~\ref{tab:parameters}.
Each secondary device is equipped with two antennas.
We place the two secondary devices at Location~1 in Fig.~\ref{fig:floor_plan}.

\begin{figure} 
	\centering
	\includegraphics[width=2in]{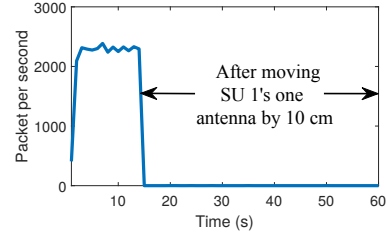}
	\caption{Packet delivery rate in the primary network before and after moving SU~1's one antenna by $10$~cm.}
	\label{fig:comerc_effect}
\end{figure}

\noindent 
\textbf{Primary Network.}
We first study the performance of the primary devices with and without the secondary devices.
Fig.~\ref{fig:comerc_primary}(a) shows the measured packet delivery rate between the two primary devices without secondary devices (i.e., the secondary devices are turned off). 
Fig.~\ref{fig:comerc_primary}(b) shows the packet delivery rate in the primary network when the secondary devices conduct their transmission in Phase~II (see Fig.~\ref{fig:experiemental_cases}(b)).
It can be seen that, in both cases, the primary network achieves almost the same packet delivery rate. 
This indicates that the primary network is almost not affected by the secondary network. 

How is the interference from the secondary transmitter handled? 
Is it because of the BBF on the secondary transmitter (SU~1)?
To answer these questions, we conduct another experiment.
When both primary and secondary networks are transmitting, we move one of the secondary transmitter's antennas about 10~cm. 
Fig.~\ref{fig:comerc_effect} shows the packet delivery rate of the primary network before and after the antenna movement. 
We can see that the movement of SU~1's one antenna results in a steep drop of primary network's packet delivery rate. 
This reveals that it is the BBF on SU~1 that effectively handles the interference for PU~2.

\begin{figure}[t]
	\centering
	\begin{subfigure}[t]{0.22\textwidth}
		\centering
		\includegraphics[width=40mm]{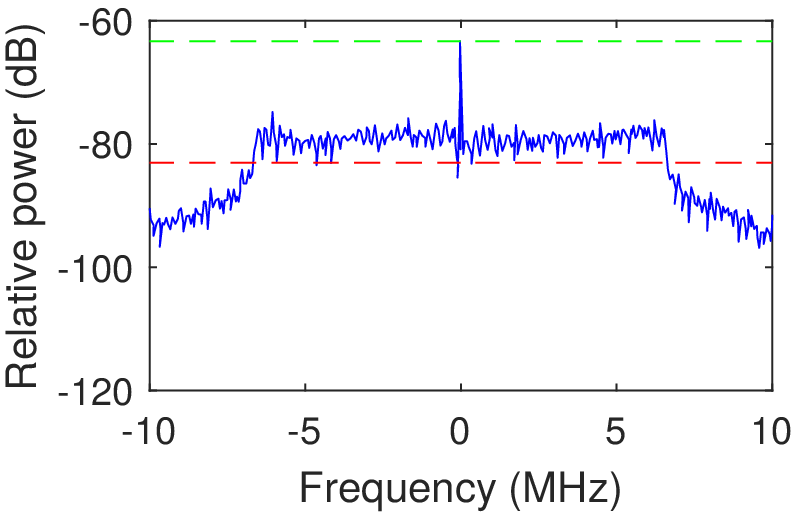}
		\caption{Received interference from PU~1 at SU~2's first antenna.}
	\end{subfigure}
	\begin{subfigure}[t]{0.21\textwidth}
		\centering
		\includegraphics[width=39mm]{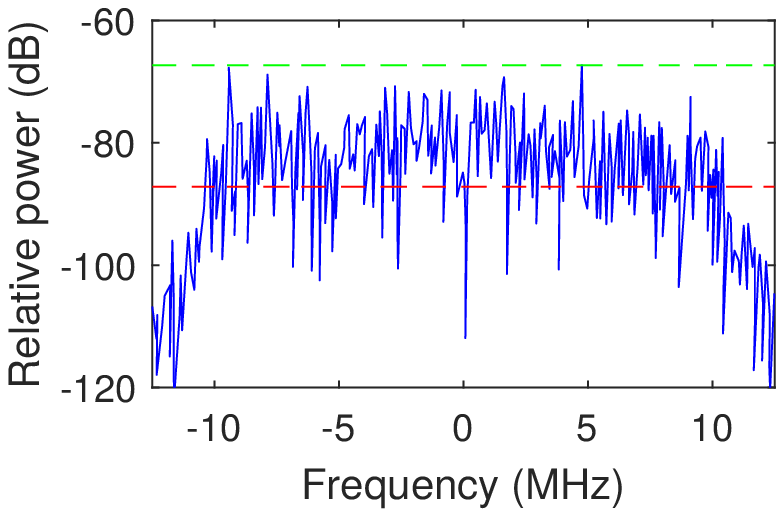}
		\caption{Received signal from SU~1 at SU~2's first antenna.}
	\end{subfigure}
	\caption{Power spectral density of the received signal and interference at the secondary receiver's first antenna.} 
	\label{fig:comerc_sig}
\end{figure}

\begin{figure}
	\centering
	\begin{subfigure}[t]{1.05in}
		\centering
		\includegraphics[width=1.1in,height=0.9in]{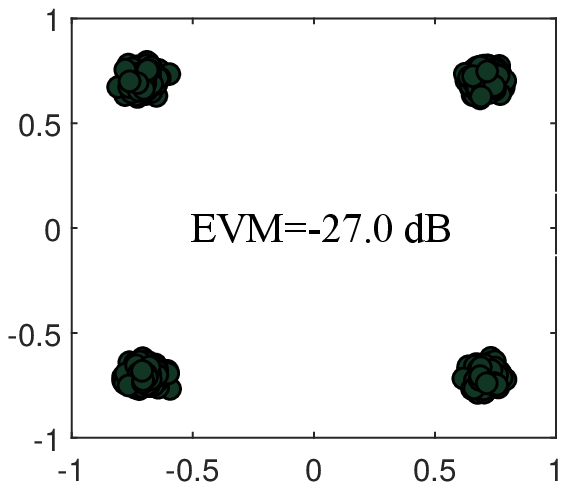}
		\caption{Interference-free scenario.}
	\end{subfigure}	
	~
	\begin{subfigure}[t]{1.05in}
		\centering
		\includegraphics[width=1.1in,height=0.9in]{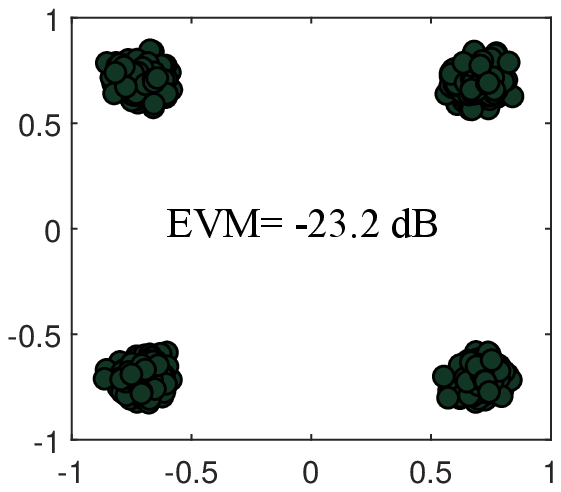}
		\caption{Spectrum sharing scenario, and SU~2 uses BIC.}
	\end{subfigure}		
	~
	\begin{subfigure}[t]{1.05in}
		\centering
		\includegraphics[width=1.1in,height=0.9in]{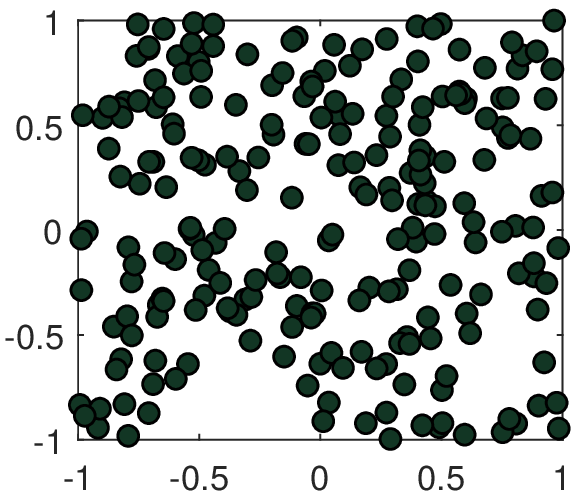}
		\caption{Spectrum sharing scenario, and SU~2 uses ZF detection.}
	\end{subfigure}
	\caption{Constellation diagram of the decoded signals at the secondary receiver (SU~2) in different scenarios.} 
	\label{fig:comerc_secondary}
\end{figure}

\noindent 
\textbf{Secondary Network.}
We now shift our focus to the secondary network.
We first check the strength of signal and interference at the secondary receiver. 
Fig.~\ref{fig:comerc_sig} shows the measured results on one of SU~2's antennas.
We can see that the signal and interference at the secondary receiver are at the similar level. This observation also holds for the another antenna. 
We then check the performance of the secondary receiver in the presence of interference from the primary transmitter. 
To do so, we conduct three experiments: 
(i) interference-free transmission of the secondary network (secondary devices only, no primary devices);
(ii) spectrum-sharing transmission with SU~2 using our proposed BIC; and 
(iii) spectrum-sharing transmission with SU~2 using ZF signal detection. 
The measured results are presented in Fig.~\ref{fig:comerc_secondary}. 
It is clear to see that, with the aid of proposed BIC, the secondary receiver can successfully decode its desired signal. 
Compared to the interference-free scenario, the EVM degradation is about 3.8~dB.
Also, the conventional ZF signal detection method is not able to decode the signal in the presence of interference. 
This shows the efficacy of our proposed BIC technique.

To show the successful coexistence of commercial Wi-Fi devices and our custom-made secondary devices, we have made a demo video and presented it in \cite{Demo2019practical}.
This video details our experimental setup and shows that the video streaming in the primary network is not harmfully affected by the concurrent data transmission of the secondary network.

\subsection{Network Setting: $(M_p=1,M_s=2)$}

We now consider a CRN where the primary devices have one antenna ($M_p = 1$) and the secondary devices have two antennas ($M_s = 2$). 
Primary networks 2 and 3 specified in Table~\ref{tab:parameters} are used in our experiments.

\subsubsection{A Case Study}

As a case study, we use primary network~3 (CDMA-like) in Table~\ref{tab:parameters} and place the secondary devices at location 1 to examine the proposed spectrum sharing scheme.

\begin{figure}
	\centering
	\begin{subfigure}[t]{0.23\textwidth}
		\centering
		\includegraphics[width=45mm]{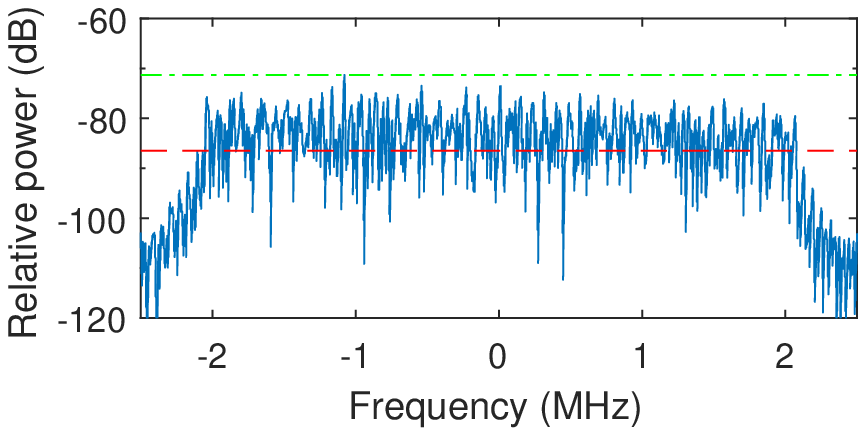}
		\caption{SU 1 uses $[\frac{1}{\sqrt{2}} ~ \frac{1}{\sqrt{2}}]$ as the precoder.}
	\end{subfigure}
	~~
	\begin{subfigure}[t]{0.23\textwidth}
		\centering
		\includegraphics[width=45mm]{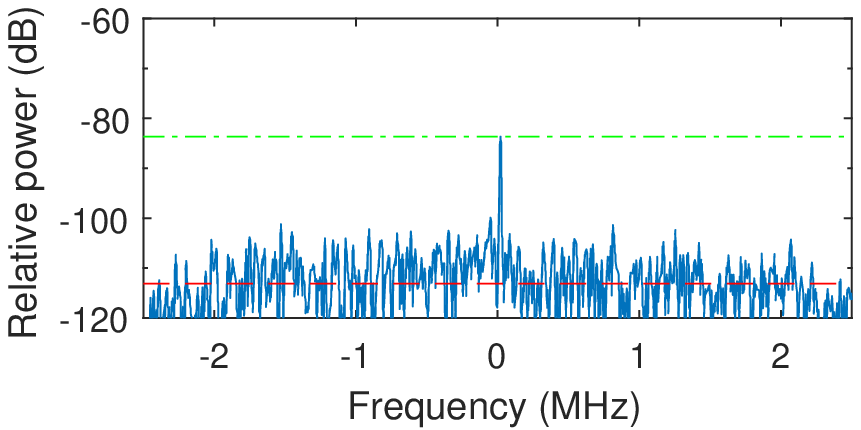}
		\caption{SU 1 uses our BBF technique.}
	\end{subfigure}
	\caption{Relative power spectral density of PU 2's received interference from SU 1 in two cases.} 
	\label{fig:tx_side_ic}
\end{figure}

\noindent
\textbf{Tx-Side IC Capability.}
We first want to quantify the Tx-side IC capability at the secondary transmitter (SU~1) from its BBF.
To do so, we conduct the following experiments.
We turn off the primary transmitter (PU 1) and measure the received interference at the primary receiver (PU 2) in two cases:
(i)~using $[\frac{1}{\sqrt{2}} ~ \frac{1}{\sqrt{2}}]$ as the precoder;
and
(ii)~using our proposed beamforming precoder in (\ref{eq:calc_eigenvectors}) and (\ref{eq:optimal_precoder}) with $\alpha_1 = 1$.
Fig.~\ref{fig:tx_side_ic} presents our experimental results.
We can see that, in the first case, the relative power spectral density of PU~2's received interference is about $-87$ dB.
In the second case, the relative power spectral density of PU~2's received interference is about $-113$ dB.
Comparing these two cases, we can see that the tx-side IC capability from BBF is about $113-87=26$~dB.  
We note that, based on our observations, the relative power spectral density of the noise at PU~2 is in the range of  $-120$~dB to $-110$~dB. 
Therefore, thanks to BBF, the interference from the secondary transmitter to the primary receiver is at the noise level.

\begin{figure}
	\centering
	\begin{subfigure}[t]{0.23\textwidth}
		\centering
		\includegraphics[width=45mm]{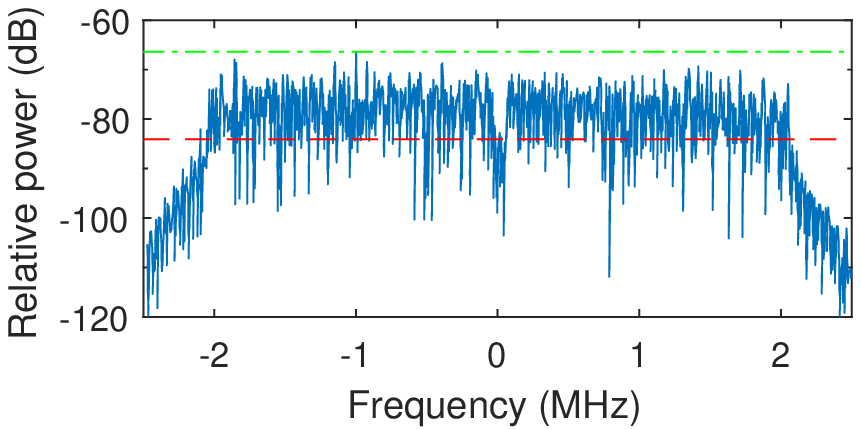}
		\caption{SU~2's received signal on its first antenna.}
	\end{subfigure}	
	~~
	\begin{subfigure}[t]{0.23\textwidth}
		\centering
		\includegraphics[width=45mm]{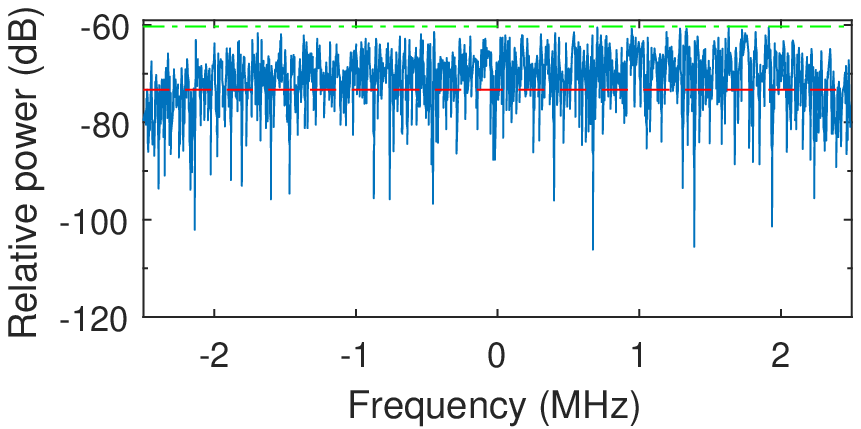}
		\caption{SU~2's received interference on its first antenna.}
	\end{subfigure}		
	\caption{Relative power spectral density of SU 2's received signal and interference on its first antenna.} 
	\label{fig:rx_sir}
\end{figure}

\noindent
\textbf{Rx-Side IC Capability, EVM, and Data Rate.}
We now study the performance of the secondary receiver (SU 2).
Firstly, we measure the signal-to-interference-ratio (SIR) at SU 2. 
Fig.~\ref{fig:rx_sir} shows our measured results on SU 2's first antenna.
We can see that the relative power spectral density of its received signal and interference is $-83$ dB and $-73$ dB, respectively. 
This indicates that the SIR on SU 2's first antenna is $-10$~dB (assuming that noise is negligible).
Using the same method, we measured that the SIR on SU 2's second antenna is $-12$~dB.

\begin{figure}
	\centering
	\begin{subfigure}[t]{0.23\textwidth}
		\centering
		\includegraphics[width=1.4in,height=1.2in]{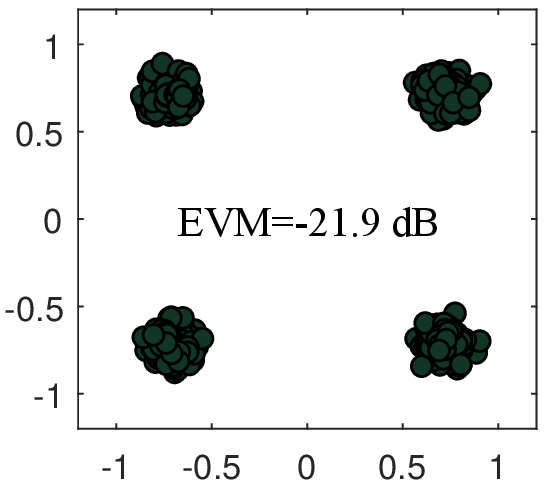}
		\caption{Decoded QPSK signals in spectrum sharing scenario.}
	\end{subfigure}
	~~
	\begin{subfigure}[t]{0.23\textwidth}
		\centering
		\includegraphics[width=1.4in,height=1.2in]{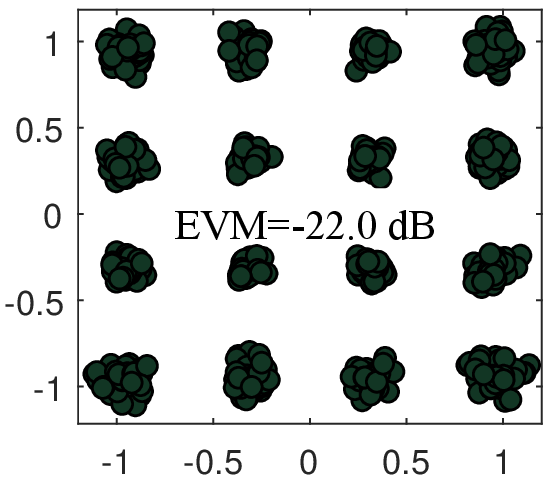}
		\caption{Decoded 16QAM signals in spectrum sharing scenario.}
	\end{subfigure}
	~~
	\begin{subfigure}[t]{0.23\textwidth}
		\centering
		\includegraphics[width=1.4in,height=1.2in]{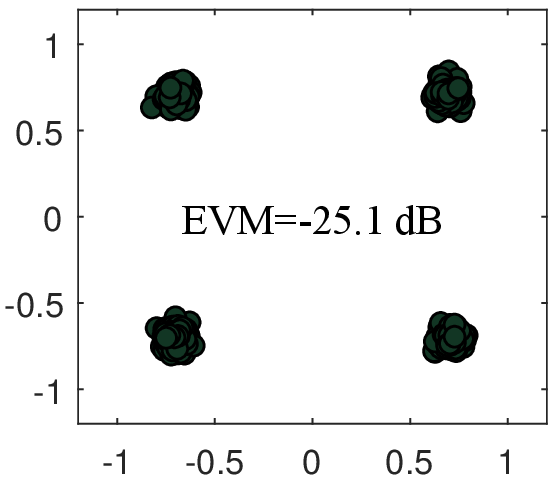}
		\caption{Decoded QPSK signals in interference-free scenario.}
	\end{subfigure}	
	~~
	\begin{subfigure}[t]{0.23\textwidth}
		\centering
		\includegraphics[width=1.4in,height=1.2in]{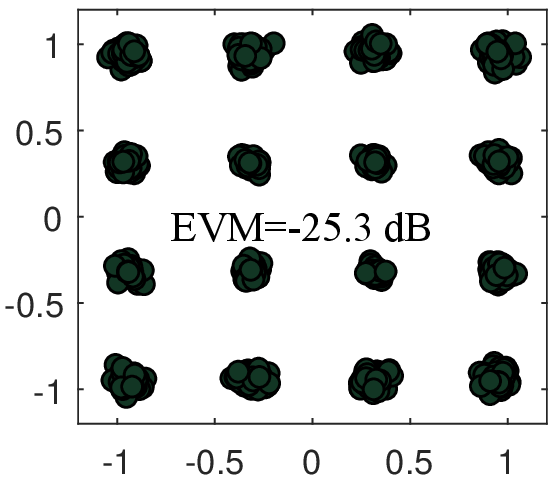}
		\caption{Decoded 16QAM signals in interference-free scenario.}
	\end{subfigure}		
	\caption{Constellation diagram of decoded signals at SU 2 in two scenarios: our proposed spectrum sharing scheme versus interference-free scenario.} 
	\label{fig:constellation}
\end{figure}

Secondly, we measure the EVM of SU 2's decoded signals in the presence of interference. 
Fig.~\ref{fig:constellation}(a--b) present the constellation of the decoded signals at SU 2. 
It is evident that SU 2 can decode both QPSK and 16QAM signals from SU~1 in the presence of interference from PU 1. 
The EVM is $-21.9$~dB when QPSK is used for the secondary network and $-22$ dB when 16QAM is used for the secondary network.
As a benchmark, Fig.~\ref{fig:constellation}(c--d) present the experimental results when there is no interference from PU~1. 
Comparing Fig.~\ref{fig:constellation}(a--b) with Fig.~\ref{fig:constellation}(c--d), we can see that SU 2 can effectively cancel the interference from PU~1.

Finally, we calculate SU 2's IC capability and throughput. 
Based on the SIR on SU 2's antennas and the EVM of its decoded signals, SU 2's IC capability is
$10+21.9=31.9$~dB in this case.
Based on (\ref{eq:evm_tp}) and the measured EVM, the throughput (data raete) of the secondary network is extrapolated to be $4.5$ Mbps.

\subsubsection{Experimental Results at all Locations}
We now extend our experiments from one location to all 12 locations and present the measured results as follows.

\begin{figure}
	\centering
	\begin{subfigure}[t]{0.23\textwidth}
		\centering
		\includegraphics[width=43mm]{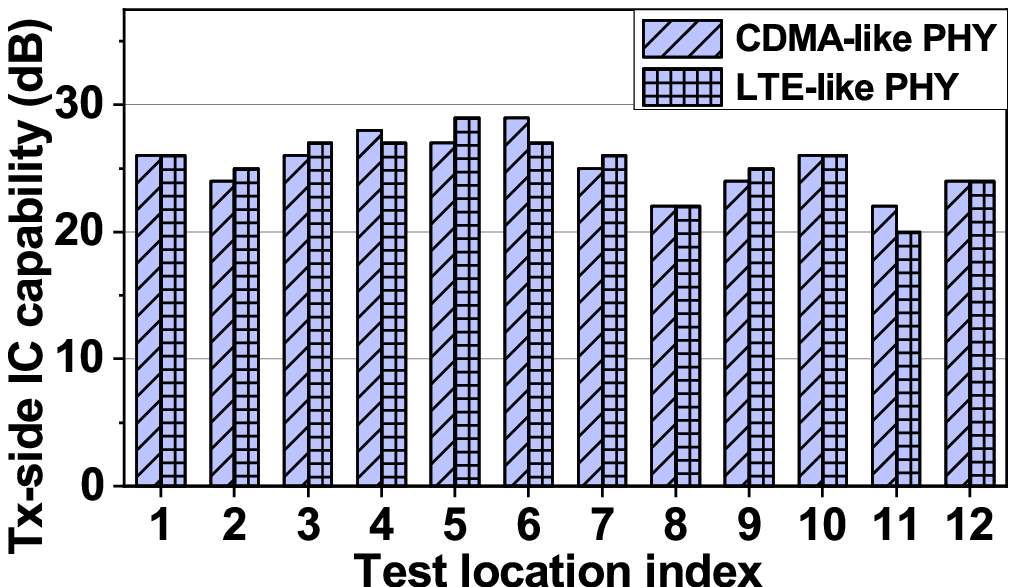}
		\caption{Tx-side IC capability from the secondary transmitter's BBF.}
	\end{subfigure}
	~~
	\begin{subfigure}[t]{0.23\textwidth}
		\centering
		\includegraphics[width=43mm]{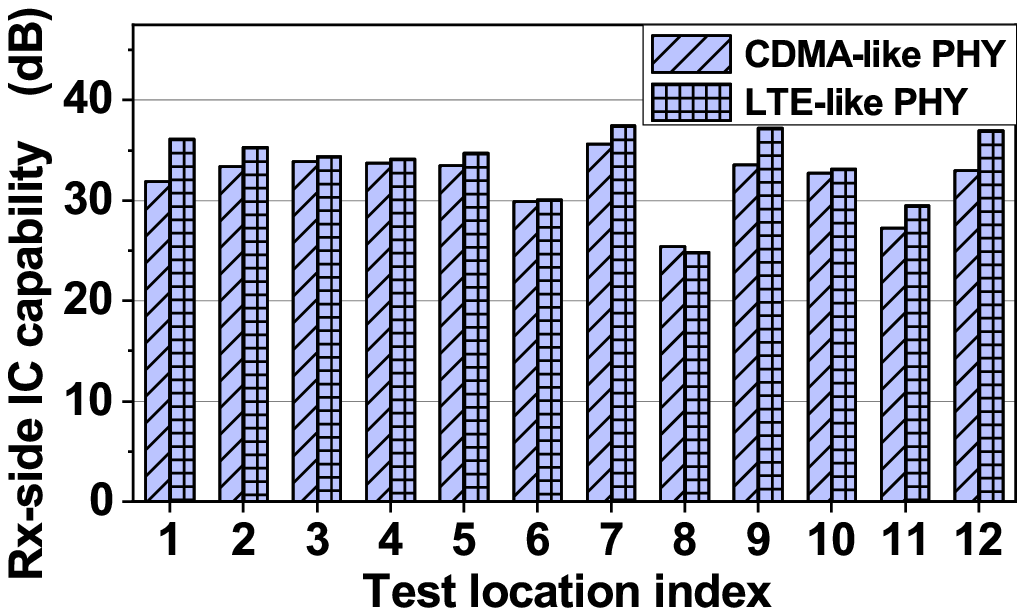}
		\caption{Rx-side IC capability from the secondary receiver's BIC.}
	\end{subfigure}
	\begin{subfigure}[t]{0.23\textwidth}
		\centering
		\includegraphics[width=43mm]{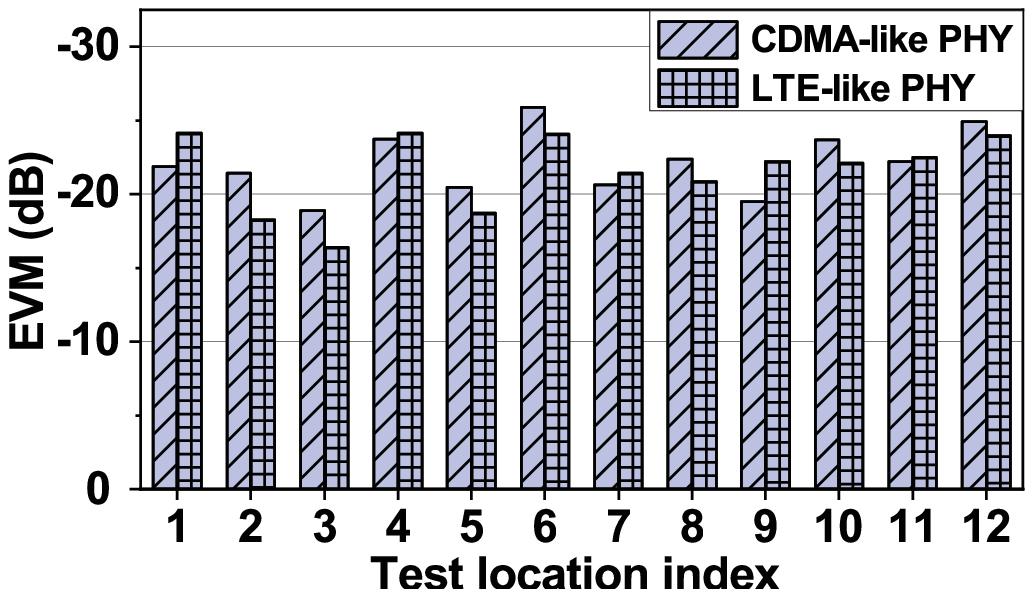}
		\caption{EVM of the decoded signals at the secondary receiver.}
	\end{subfigure}	
	~~
	\begin{subfigure}[t]{0.23\textwidth}
		\centering
		\includegraphics[width=43mm]{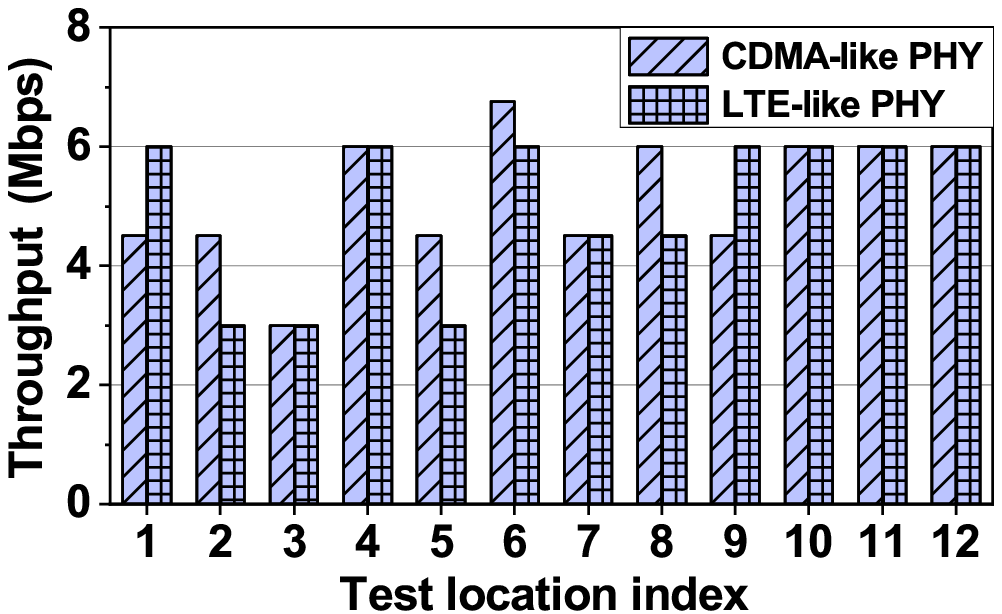}
		\caption{Throughput of the secondary network.}
	\end{subfigure}		
	\caption{Performance of the proposed spectrum sharing scheme when each secondary device has two antennas and each primary device has one antenna.} 
	\label{fig:performance}
\end{figure}

\noindent
\textbf{Tx-Side IC Capability.}
Fig.~\ref{fig:performance}(a) presents the tx-side IC capability of the two-antenna secondary transmitter (SU~1).
We can see that the secondary transmitter achieves a minimum of $20.0$~dB and an average of $25.3$~dB IC capability in the 12~locations.

\noindent
\textbf{Rx-Side IC Capability.}
Fig.~\ref{fig:performance}(b) presents the rx-side IC capability of the two-antenna secondary receiver.
We can see that the secondary receiver achieves a minimum of $25.0$~dB, a maximum of $38.0$~dB, and an average of $32.8$~dB IC capability in the 12 locations, regardless of the PHY used for the primary network.

\noindent
\textbf{Rx-Side EVM.}
Fig.~\ref{fig:performance}(c) presents the EVM of the decoded signals at the two-antenna secondary receiver in the presence of interference from the primary transmitter.
We can see that in all the locations, although the EVM varies, the EVM achieves a maximum of $-16.4$ dB, a minimum of $-25.9$~dB, an average of $-21.8$ dB in the 12 locations, regardless of the PHY used for the primary network.

\noindent
\textbf{Throughput of Secondary Network.}
Based on the measured EVM at the secondary receiver, we extrapolate the achievable data rate in the secondary network using \eqref{eq:evm_tp}.
Fig.~\ref{fig:performance}(d) presents the results.
As we can see, the secondary network achieves a minimum of $3.0$~Mbps data rate, a maximum of $6.7$~Mbps, and an average of $5.1$~Mbps in the 12 locations. 
Note that this data rate is achieved by the secondary network in $5$~MHz bandwidth, and the secondary transmitter's power is controlled so that its interference at the primary receiver (after BBF) remains at the noise level.

\subsubsection{BBF versus Other Beamforming Techniques}

As BBF is the core component of our spectrum sharing scheme, we would like to further examine its performance by comparing it against the following two beamforming techniques.

\begin{itemize}

\item 
\textit{Explicit Beamforming (EBF)}: 
In this technique, the secondary transmitter (SU~1) has the \textit{forward} channel knowledge between itself and the primary receiver (PU~2), i.e., $\mathbf{H}^{\left[1\right]}_{\text{sp}}(k)$. 
The forward channel knowledge is obtained through explicit channel feedback.
Specifically, SU~1 sends a null data packet (NDP) to PU~2, which estimates the channel and feed the estimated channel information back to SU~1.
After obtaining the forward channel $\mathbf{H}^{\left[1\right]}_{\text{sp}}(k)$, SU~1 constructs the precoder by
$\mathbf{P}(k) = \mathit{mineigvectors}(\mathbf{H}^{\left[1\right]}_{\text{sp}}(k))$, 
where $k$ is subcarrier index.

\item 
\textit{Implicit Beamforming (IBF)}: 
In this technique, the secondary transmitter (SU~1) has the \textit{backward} channel knowledge from the primary receiver (PU~2) to itself, i.e., $\mathbf{H}^{\left[1\right]}_{\text{ps}}(k)$. 
The backward channel knowledge is obtained through implicit channel feedback.
Specifically, PU~2 sends a null data packet (NDP) to SU~1.
SU~1 first estimates the backward channel $\mathbf{H}^{\left[1\right]}_{\text{ps}}(k)$.
It then constructs the precoder by
$\mathbf{P}(k) = \mathit{mineigvectors}(\mathbf{H}^{\left[1\right]}_{\text{ps}}(k))$, 
where $k$ is subcarrier index.
Channel calibration has been performed at SU~1 before signal transmission.

\end{itemize}

\begin{figure}
	\centering
	\includegraphics[width=90mm]{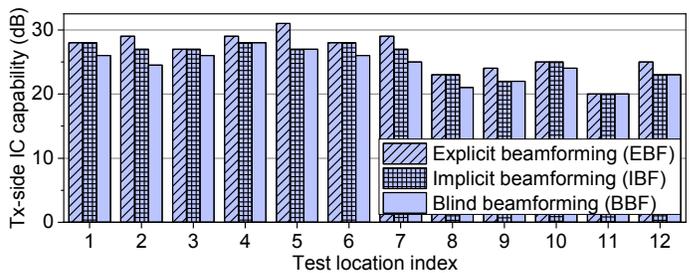}
	\caption{Comparison of tx-side IC capability of the three beamforming techniques.}
	\label{fig:tx_IC_bounds}
\end{figure}

We conduct experiments to measure the tx-side IC capability of these three beamforming techniques.
Fig.~\ref{fig:tx_IC_bounds} depicts our results.
We can see that, compared to EBF, our proposed BBF has a maximum of 4.5~dB and an average of 2.1~dB degradation. 
Compared to IBF, our proposed BBF has a maximum of 2.5~dB and an average of 1.0~dB degradation. 
The results show that the proposed BBF has competitive performance compared to EBF and IBF. 
We note that, although offering better performance, EBF and IBF cannot be used in underlay CRNs as they require knowledge and cooperation from the primary devices.

%

\begin{figure}
	\centering
	\begin{subfigure}[t]{0.23\textwidth}
		\centering
		\includegraphics[width=43mm]{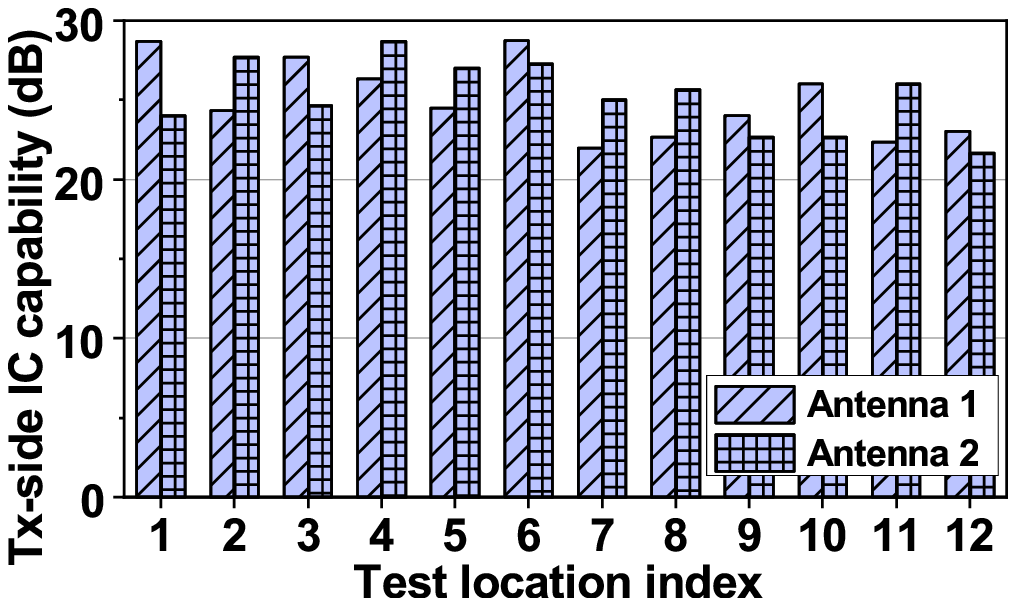}
		\caption{Tx-side IC capability from the secondary transmitter's BBF.}
	\end{subfigure}	
	~
	\begin{subfigure}[t]{0.23\textwidth}
		\centering
		\includegraphics[width=43mm]{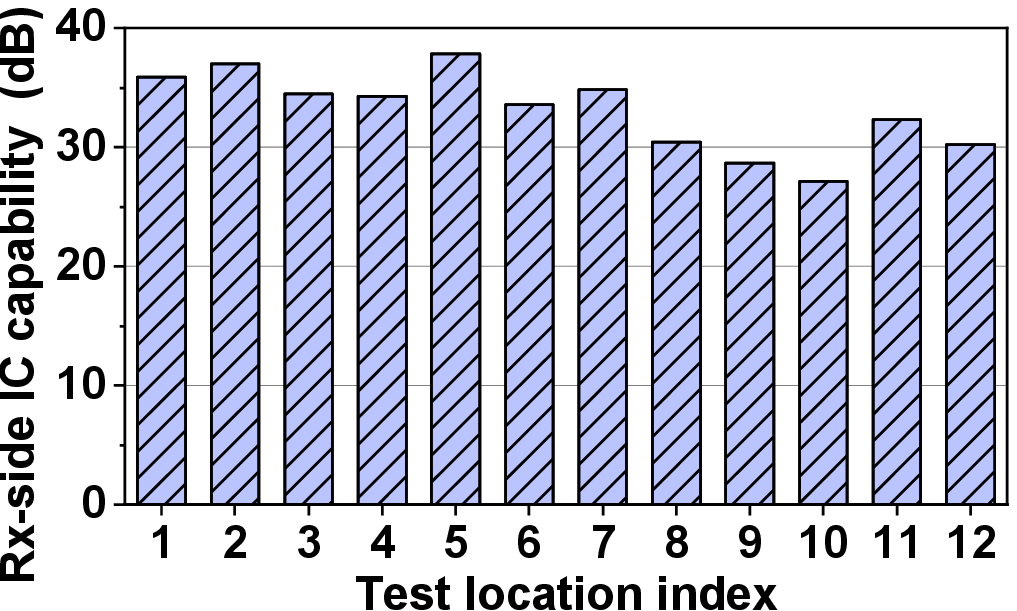}
		\caption{Rx-side IC capability from the secondary receiver's BIC.}
	\end{subfigure}	
	\caption{Tx-side and rx-side IC capabilities of a secondary network where each device has three antennas.} 
	\label{fig:ic_capability_case2x3}
\end{figure}

\begin{figure}
	\begin{subfigure}[t]{0.23\textwidth}
		\centering
		\includegraphics[width=43mm]{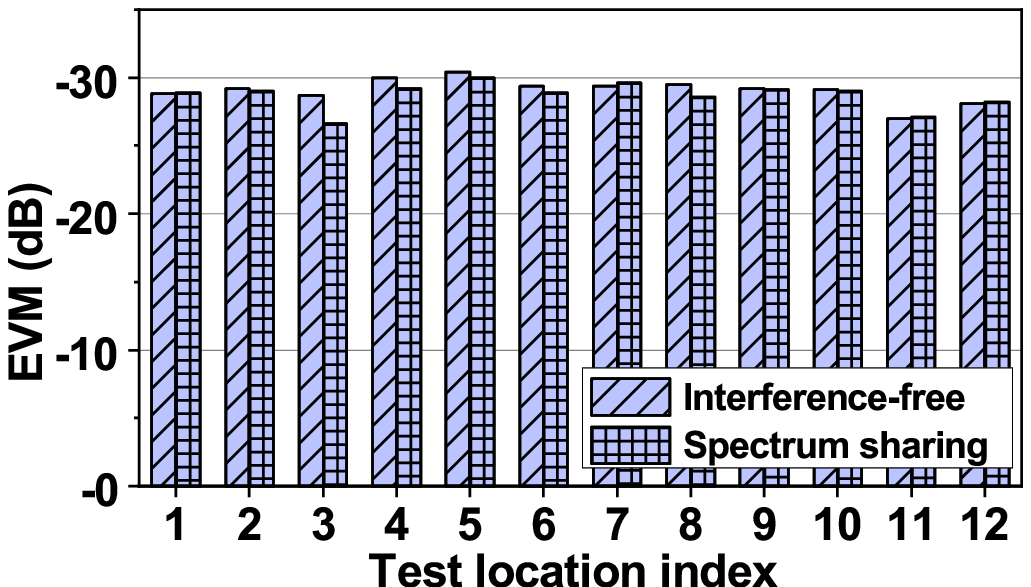}
		\caption{EVM of the decoded data stream~1 at the primary receiver.}
	\end{subfigure}
	~
	\begin{subfigure}[t]{0.23\textwidth}
		\centering
		\includegraphics[width=43mm]{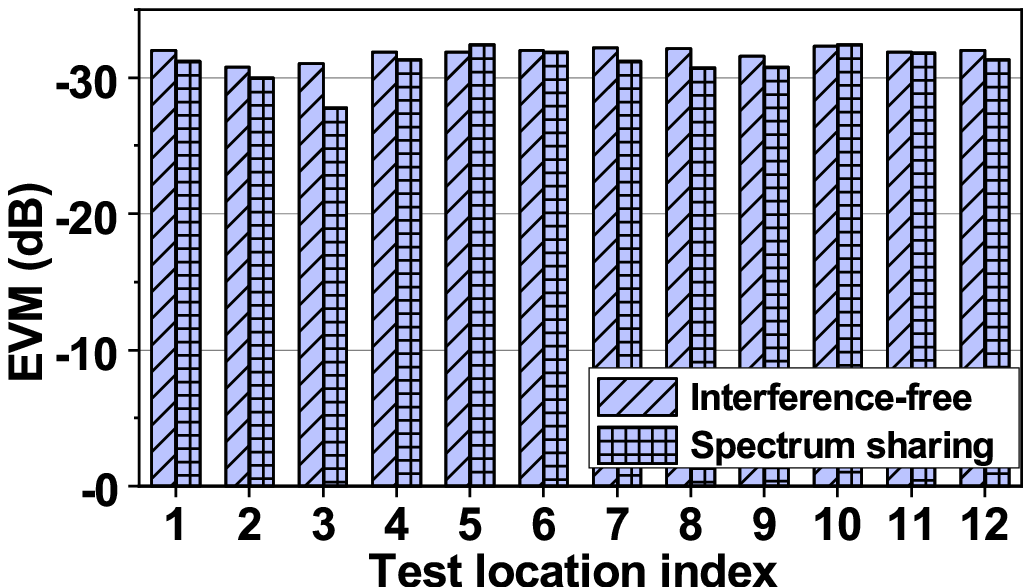}
		\caption{EVM of the decoded data stream~2 at the primary receiver.}
	\end{subfigure}
	\caption{EVM performance of the two data streams in the primary network with and without the secondary network.}
	\label{fig:evm_primary_case2x3}
\end{figure}

\subsection{Network Setting: $(M_p=2,M_s=3)$}

In this subsection, we study the CRN in Fig.~\ref{fig:experiemental_cases} where the primary devices have two antennas and the secondary devices have three antennas (i.e., $M_p = 2$ and $M_s = 3$). 
The primary devices use their two antennas for spatial multiplexing.
That is, two independent data streams are transfered in the primary network. 
The secondary devices use their spatial DoF provided by their three antennas for both interference management and signal transmission.
Indeed, one data stream is transfered in the secondary network.
The primary network uses LTE-like PHY (see primary network 2 in Table~\ref{tab:parameters}) for data transmission. 
We study our spectrum sharing scheme in this CRN and report the measured results below.

\noindent
\textbf{Tx-Side IC Capability:} 
In this CRN, since the primary receiver has two antennas, the secondary transmitter needs to cancel its generated interference for both antennas on the primary receiver. 
We measure the IC capability of our proposed BBF for the primary receiver's both antennas. 
Fig~\ref{fig:ic_capability_case2x3}(a) exhibits our measured results. 
We can see that a three-antenna secondary transmitter can effectively cancel the interference on the primary receiver's both antennas. 
Specifically, the BBF on the secondary transmitter achieves a minimum of $21.7$~dB, a maximum of $28.7$~dB, and an average of $25.1$~dB IC capability for the primary receiver's two antennas.

\noindent
\textbf{Rx-Side IC Capability:} 
In this CRN, since the primary transmitter sends two independent data streams, the secondary receiver needs to decode its desired signals in the presence of two interference sources. 
We measure the rx-side IC capability of our proposed BIC at the three-antenna secondary receiver. 
Fig~\ref{fig:ic_capability_case2x3}(b) exhibits our measured results.
We can see that the prosed BIC on the secondary receiver achieves a minimum of $26.5$~dB, a maximum of $38.1$~dB, and an average of $33.0$~dB IC capability over the 12 locations. 
This shows the effectiveness of the proposed BIC in handling unknown interference.

\noindent
\textbf{EVM at Primary Receiver:} 
We now study the performance of the two data streams in the primary network.
We want to see if the presence of secondary network harmfully affects the traffic in the primary network. 
To do so, we measure the EVM of the decoded two data streams at the primary receiver in two cases:
(i) in the presence of the secondary network and (ii) in the absence of the secondary network. 
Fig.~\ref{fig:evm_primary_case2x3} presents our measured results. 
It can be seen that the presence of the secondary network does not visibly affect the EVM performance of the primary network. 
This indicates that the BBF at the secondary network successfully handles the interference from the secondary transmitter to the primary receiver.

\begin{figure}
	\begin{subfigure}[t]{0.23\textwidth}
		\centering
		\includegraphics[width=43mm]{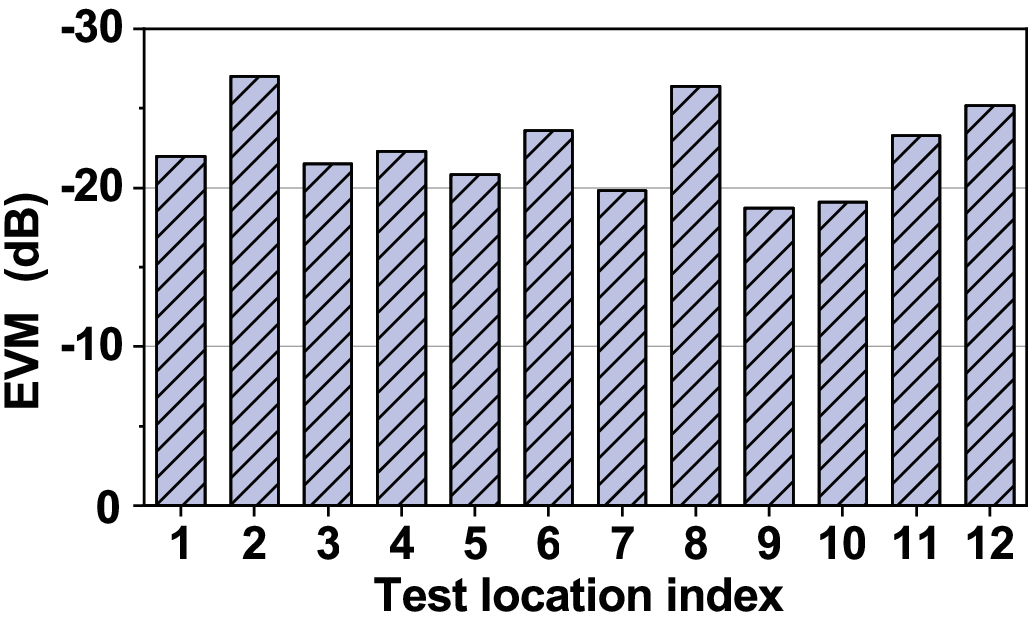}
		\caption{EVM of decoded signals at the secondary receiver.}
	\end{subfigure}
	~~
	\begin{subfigure}[t]{0.23\textwidth}
		\centering
		\includegraphics[width=43mm]{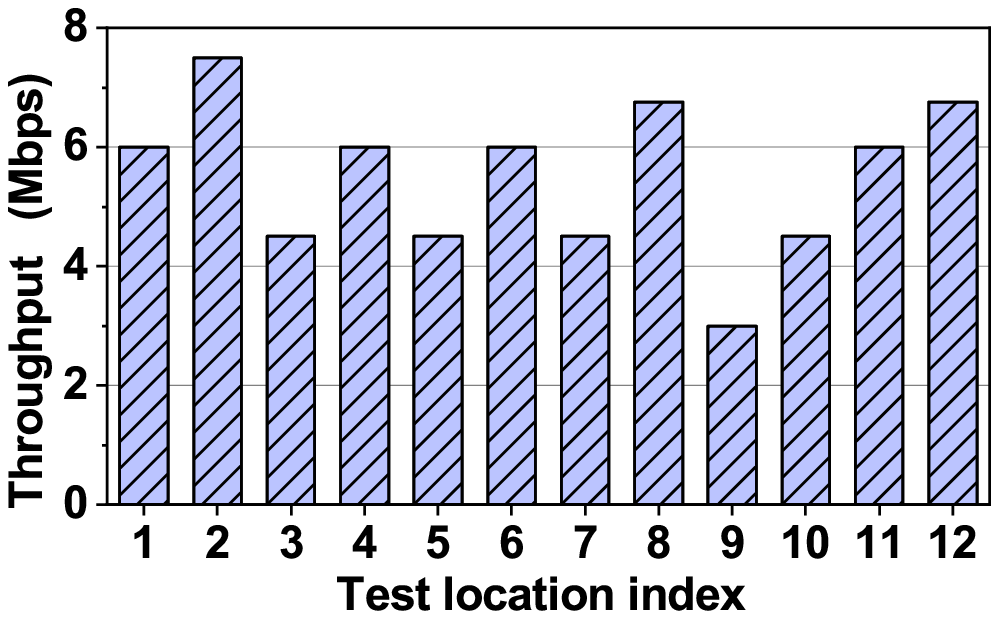}
		\caption{Throughput of the secondary network.}
	\end{subfigure}	
	\caption{Performance of the secondary network in the proposed spectrum sharing scheme.} 
	\label{fig:perf_case2x3}
\end{figure}

\noindent
\textbf{EVM at Secondary Receiver:} 
Having confirmed that the spectrum utilization of secondary network does not degrade the performance of primary network, we now study the achievable throughput of the secondary network. 
Recall that we transfer one data stream in the secondary network. 
We measure the EVM performance of the decoded signal at the secondary receiver. 
Fig.~\ref{fig:perf_case2x3}(a) depicts the measured results. 
We can see that the EVM at the secondary receiver achieves a minimum of $-27.7$~dB, a maximum of $-18.2$~dB, and an average of $-22.5$~dB over the 12 locations.

\noindent
\textbf{Throughput of Secondary Network:} 
Based on the measured EVM at the secondary receiver, we extrapolate the achievable data rate of the secondary network using \eqref{eq:evm_tp}.
The extrapolated data rate is presented in Fig.~\ref{fig:perf_case2x3}(b).
We can see that the proposed spectrum sharing scheme achieves a minimum of $3.0$~Mbps, a maximum of $7.5$~Mbps, and an average of $5.5$~Mbps over the 12 locations. 
Note that this data rate is achieved by the secondary network in $5$~MHz and without harmfully affecting the primary network.

\subsection{Summary of Observations}

\begin{figure} 
	\centering
	\includegraphics[width=85mm,height=40mm]{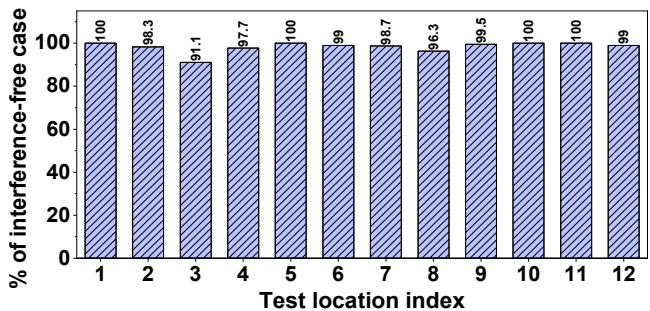}
	\caption{Performance evaluation of the proposed spectrum sharing scheme at a glance.}
	\label{fig:primary_network_summary}
\end{figure}

We now summarize the observations from our experimental results as follows:

\begin{itemize}
	\item 
	\textit{BBF:}
	BBF demonstrates its capability of handling cross-network interference in CRNs where the secondary network has no knowledge about the primary network. 
	In $(M_p = 1, M_s = 2)$ network setting, BBF achieves an average of $25.3$~dB IC capability. 
	In $(M_p = 2, M_s = 3)$ network setting, BBF achieves an average of $25.1$~dB IC capability. 
	
	\item 
	\textit{BIC:}
	BIC also demonstrates its capability of decoding its desired signal in the presence of unknown interference. 
	In $(M_p = 1, M_s = 2)$ network setting, it achieves an average of $32.8$~dB IC capability. 
	In $(M_p = 2, M_s = 3)$ network setting, it achieves an average of $33.0$~dB IC capability. 
	
	\item 
	\textit{Primary Network:}
	For the CRN with both network settings, the primary network has very small performance degradation when the secondary network shares the spectrum (compared to the case without secondary network).
	As shown in Fig.~\ref{fig:primary_network_summary}, the average degradation of EVM performance at the primary receiver is $1.7$\% over the 12~locations. 
	
	\item 
	\textit{Secondary Network:}		
	Using BBF at its transmitter and BIC at its receiver, the secondary network intends to establish communications by sharing the spectrum with the primary network.
	The secondary network achieves $1.0$~bits/s/Hz in the CRN with network setting $(M_p = 1, M_s = 2)$ and $1.1$~bits/s/Hz in the CRN with network setting $(M_p = 2, M_s = 3)$.
	
\end{itemize}


\section{Limitations and Discussions}
\label{sec:limitation}

While the proposed scheme demonstrates its potential in real-world network settings, there are still some issues that remain open and need to be addressed prior to its real-world applications.

\noindent
\textbf{Primary Traffic Directions.}
In our spectrum sharing scheme, we assume that the primary communications are bi-directional and that the pattern of primary traffic is consistent. 
Under such assumptions, the secondary devices are easy to learn the direction (forward or backward) of primary traffic and therefore overhear the backward interfering signals for the construction of forward beamforming filters. 
In real systems, the pattern of primary traffic is not consistent necessarily. 
A forward data packet may not be followed by an ACK/NACK packet. 
In such network scenarios, a sophisticated learning algorithm is needed for the secondary devices to differentiate the forward and backward transmissions in the primary network. 
Signal signatures such as relative-signal-strength (RSS) and angle-of-arrival (AoA) can be utilized for the design of such a learning algorithm.

\noindent
\textbf{Channel Coherence Time.}
As wireless channels vary over time, the constructed beamforming filters at the secondary transmitter are valid only within the period of channel coherence time. 
In static networks (e.g., indoor Wi-Fi), the devices are stationary or moving at a low speed. 
The channel coherence time is large enough to cover the entire period of primary forward transmission and, as a result, the secondary network can use the spectrum during the entire period of primary forward transmission.
But in the dynamic networks with highly mobile devices, the channel coherence time may be smaller than the duration of primary forward transmission. 
In such a case, the secondary network cannot use the entire airtime of primary forward transmission. 
Instead, it can only access the spectrum when its beamforming filters remain valid (i.e., within the channel coherence time).

\noindent
\textbf{Half-Time Spectrum Utilization:}
For our spectrum sharing scheme, Fig.~\ref{fig:mac_protocol} illustrates a half-time spectrum utilization of the secondary network. 
We note that this is neither an upper bound nor a lower bound of the airtime utilization in the secondary network.
When the channel coherence time is sufficiently large, the constructed beamforming filters remain valid for a long time. 
In such a case, the airtime utilization of the secondary network can approach to $100$\%. 
But when the channel coherence time is small, the airtime utilization of the secondary network could be very low (approaching to zero).

\noindent
\textbf{Ill-Conditioned MIMO Channel:}
In practice, ill-conditioning of a MIMO channel can be attributed to high correlation of transmit antennas, high correlation of receive antennas, or both. 
Both BBF and BIC techniques rely on the assumption that the effective spatial DoF at a secondary device is more than that of a primary device.
This means that the BBF and BIC techniques are resilient to the correlation of the primary devices' antennas, but susceptible to the correlation of the secondary devices' antennas. 
When a secondary device lacks spatial DoF, a way to overcome this issue is to increase the number of its physical antennas. 
Therefore, massive MIMO permits a huge potential of our spectrum sharing scheme.

\noindent
\textbf{Large-Scale CRNs:}
In this paper, we consider a small CRN that comprises a pair of primary users and a pair of secondary users. 
Extending the proposed spectrum sharing scheme to a large-scale CRN requires a holistic protocol that can fully exploit BBF and BIC at the secondary devices.

\section{Conclusion}
\label{sec:conclusion}
In this paper, we proposed a spectrum sharing scheme for an underlay CRN that comprises two primary users and two secondary users.
The proposed scheme allows the secondary users to use the spectrum without harmfully affecting the performance of the primary users.
The key components of our scheme are two MIMO-based IC techniques: BBF and BIC.
BBF enables the secondary transmitter to pre-cancel its generated interference for the primary receiver.
BIC enables the secondary receiver to decode its desired signal in the presence of unknown interference from the primary transmitter. 
Collectively, these two IC techniques make it possible for the secondary users to access the spectrum while remaining transparent to the primary users. 
We have built a prototype of our spectrum sharing scheme on a GNURadio-USRP2 wireless testbed and demonstrated that our prototyped secondary devices can coexist with commercial Wi-Fi devices.
Experimental results further show that, for a secondary user with two or three antennas, BBF and BIC achieve about 25~dB and 33~dB IC capability in an office environment, respectively.

\begin{appendices}
	
\section{Proof of Lemma~\ref{lem:bbf}}
\label{app:bbf}

We first consider the signal transmission in Phase~I and then consider that in Phase~II.
In Phase~I, if the noise is zero, we have 
$\mathbf{Y}(l,k) = \mathbf{H}_\mathrm{sp}^\mathrm{[1]}(k) \mathbf{X}_\mathrm{p}^\mathrm{[1]}(l,k)$.
Then, we have 
\begin{align}
\sum_{l=1}^{L_\mathrm{p}}\mathbf{Y}(l,k) \mathbf{Y}(l,k)^*
&\overset{(a)}{=}
L_\mathrm{p} \mathbb{E}[\mathbf{Y}(l,k) \mathbf{Y}(l,k)^*] 
\nonumber\\
&\overset{(b)}{=}
L_\mathrm{p} \mathbf{H}_\mathrm{sp}^\mathrm{[1]}(k) \mathbf{R}_\mathrm{x}(k) \mathbf{H}_\mathrm{sp}^\mathrm{[1]}(k)^*,
\label{eq:Y_deductoin}
\end{align}
where (a) follows from that $\mathbf{Y}(l,k)$ is a stationary random process, which is true in practice; and
(b) follows from the definition of $\mathbf{R}_\mathrm{x}(k) = \mathbb{E}[\mathbf{X}_\mathrm{p}^\mathrm{[1]}(l,k) \mathbf{X}_\mathrm{p}^\mathrm{[1]}(l,k)^*]$.

Based on \eqref{eq:Y_deductoin}, we have
\begin{align}
\!
\mathit{Rank}\Big(\!\!\sum_{l=1}^{L_\mathrm{p}}\mathbf{Y}(l,k) \mathbf{Y}(l,k)^*\!\Big)
& \!\!=\!\!
\mathit{Rank}\Big(\!L_\mathrm{p} \mathbf{H}_\mathrm{sp}^\mathrm{[1]}(k) \mathbf{R}_\mathrm{x}(k) \mathbf{H}_\mathrm{sp}^\mathrm{[1]}(k)^*\!\Big)
\nonumber \\
& \!\le 
\mathit{Rank}\Big(\mathbf{R}_\mathrm{x}(k)\Big)
\le 
M_\mathrm{p}.
\label{eq:rank}
\end{align}

Inequation (\ref{eq:rank}) indicates that $\sum_{l=1}^{L_\mathrm{p}}\mathbf{Y}(l,k) \mathbf{Y}(l,k)^*$ has at least $M_\mathrm{s} - M_\mathrm{p}$ eigenvectors that correspond to zero eigenvalues.
This further indicates that $[\mathbf{U}_1, \mathbf{U}_2, \cdots, \mathbf{U}_{M_\mathrm{e}}]$ in (\ref{eq:calc_eigenvectors}) are corresponding to zero eigenvalues. 
Therefore, we have 
\begin{equation}
\Big(\sum_{l=1}^{L_\mathrm{p}}\mathbf{Y}(l,k) \mathbf{Y}(l,k)^*\Big) \mathbf{U}_m = \mathbf{0},
~~
\mbox{for $1 \le m \le M_\mathrm{e}$}.
\label{eq:a}
\end{equation}

Based on (\ref{eq:Y_deductoin}) and (\ref{eq:a}), we have 
\begin{equation}
\Big(L_\mathrm{p} \mathbf{H}_\mathrm{sp}^\mathrm{[1]}(k) \mathbf{R}_\mathrm{x}(k) \mathbf{H}_\mathrm{sp}^\mathrm{[1]}(k)^* \Big) \mathbf{U}_m = \mathbf{0},
~~
\mbox{for $1 \le m \le M_\mathrm{e}$}.
\label{eq:b}
\end{equation}

In practical wireless environments,  we have 
$\mathit{Rank}\big(\mathbf{H}_\mathrm{sp}^\mathrm{[1]}(k) \big) = M_\mathrm{p}$
and
$\mathit{Rank}\big(\mathbf{R}_\mathrm{x}(k)\big) = M_\mathrm{p}$.
Therefore, the following equation can be deducted from (\ref{eq:b}).
\begin{equation}
\mathbf{H}_\mathrm{sp}^\mathrm{[1]}(k)^* \mathbf{U}_m = \mathbf{0},
~~
\mbox{for $1 \le m \le M_\mathrm{e}$}.
\label{eq:c}
\end{equation}

Based on (\ref{eq:optimal_precoder}) and (\ref{eq:c}), we have 
\begin{equation}
\mathbf{H}_\mathrm{sp}^\mathrm{[1]}(k)^*  \mathbf{P}(k)
=
\sum_{m = 1}^{M_\mathrm{e}} \alpha_m  \mathbf{H}_\mathrm{sp}^\mathrm{[1]}(k)^* \mathbf{U}_m 
= 
\mathbf{0}.
\label{eq:d}
\end{equation}

We now consider signal transmission in Phase~II (see Fig.~\ref{fig:spectrum_sharing_scheme}(b)).
Denote $\mathbf{H}_\mathrm{ps}^\mathrm{[2]}$ as the matrix representation of the channel from SU~1 to PU~2 on subcarrier $k$ in Phase~II.
Given that the forward and backward channels in the two phases are reciprocal, we have 
$\mathbf{H}_\mathrm{ps}^\mathrm{[2]} = \big({\mathbf{H}_\mathrm{sp}^\mathrm{[1]}}\big)^T$.
Then, we have 
\begin{equation}
\mathbf{H}_\mathrm{ps}^\mathrm{[2]}(k) \overline{\mathbf{P}(k)} 
= 
\big({\mathbf{H}_\mathrm{sp}^\mathrm{[1]}}\big)^T \overline{\mathbf{P}(k)} 
=
\overline{\mathbf{H}_\mathrm{sp}^\mathrm{[1]}(k)^* \mathbf{P}(k)} 
= 
\mathbf{0}.
\end{equation}

It means that the precoding vector $\overline{\mathbf{P}(k)}$ is orthogonal to the interference channel $\mathbf{H}_\mathrm{ps}^\mathrm{[2]}(k)$.
Therefore, we conclude that the proposed beamforming scheme can completely pre-cancel the interference from the secondary transmitter at the primary receiver in Phase II.

\section{Proof of Lemma~\ref{lem:bic}}
\label{app:bic}

For notational simplicity, we denote $\mathbf{H}(k)$ as the compound channel between the SU~2 and the two transmitters (SU~1 and PU~1), 
i.e., $\mathbf{H}(k)  = \Big[\mathbf{H}_\mathrm{ss}^\mathrm{[2]}(k) \overline{\mathbf{P}(k)}  \;~ \mathbf{H}_\mathrm{sp}^\mathrm{[2]}(k) \Big]$;
we also denote $\mathbf{X}(l, k)$ as the compound transmit signals at the two transmitters, 
i.e., 
$\mathbf{X}(l, k) = \Big[X_\mathrm{s}^\mathrm{[2]}(l, k) \;~ \mathbf{X}_\mathrm{p}^\mathrm{[2]}(l, k) \Big]^T$.
Then, in noise-negligible scenarios, (\ref{eq:mimo_system_model}) can be rewritten as:
\begin{equation}
\mathbf{Y}(l, k) = 
\mathbf{H}(k) \mathbf{X}(l, k).
\label{eq:mimo_system_model_nonoise}
\end{equation}

By defining $\mathbf{R}_\mathrm{X}$ as the autocorrelation matrix of the compound transmit signals, we have 
\begin{equation}
\mathbf{R}_\mathrm{X} 
= \mathbb{E}(\mathbf{X}\mathbf{X}^H)
\overset{(a)}{=}
\left[
\begin{array}{ll}
{R}_\mathrm{xs} ~~~ \mathbf{0} \\
\mathbf{0} ~~ \mathbf{R}_\mathrm{xp} \\
\end{array}
\right]
=
\left[
\begin{array}{ll}
1 ~~~ \mathbf{0} \\
\mathbf{0} ~~ \mathbf{R}_\mathrm{xp} \\
\end{array}
\right],
\label{eq:correlation}
\end{equation}
where ${R}_\mathrm{xs}$ is the autocorrelation of SU~1's transmit signal and $\mathbf{R}_\mathrm{xp}$ is the autocorrelation matrix of PU 1's transmit signals.
(a) follows from our assumption that the transmit signal from SU~1 is independent of the transmit signals from PU~1. 
Note that $\mathbf{R}_\mathrm{xp}$ is not necessarily an identity matrix since the signals from PU~1's different antennas might be correlated.

Based on (\ref{eq:bic_filter}), (\ref{eq:mimo_system_model_nonoise}), and (\ref{eq:correlation}), we have
\begin{align}
\mathbf{G}(k)
\!
&
=\!\!
\Big[
\!\!\!\!\! 
\sum_{(l,k') \in \mathcal{Q}_k} 
\!\!\!\!\! \!
{\mathbf{Y}}(l,k') {\mathbf{Y}}(l,k')^H\Big]^{+}
\Big[
\!\!\!\!\! 
\sum_{(l,k') \in \mathcal{Q}_k} 
\!\!\!\!\! 
{\mathbf{Y}}(l,k') {X_\mathrm{s}^\mathrm{[2]}(l,k')}^*\Big] 
\nonumber \\
&  
\overset{(a)}{=}
\mathbb{E} 
\big[
{\mathbf{Y}}(l,k) {\mathbf{Y}}(l,k)^*\big]^{+}
\mathbb{E} 
\big[
{\mathbf{Y}}(l,k) {X_\mathrm{s}^\mathrm{[2]}(l,k)}^*\big] 
\nonumber \\
&  
\overset{(b)}{=}
\big[
{\mathbf{H}}(k) \mathbf{R}_\mathrm{X}  {\mathbf{H}}(k)^*\big]^{+}
\big[
{\mathbf{H}}(k) \mathbf{I}_1\big], 
\label{eq:mud_proof}
\end{align}
where (a) follows from our assumption that the amount of reference signals is sufficient to achieve convergence of $\mathbf{G}(k)$;
(b) follows from the definition that $\mathbf{I}_1$ is a vector where its first entry is 1 and all other entries are 0.

Based on (\ref{eq:decoded_signal}) and (\ref{eq:mud_proof}), we have
\begin{align}
\hat{X}_\mathrm{s}^\mathrm{[2]}(l, k) 
& =
\mathbf{G}(k)^* \mathbf{Y}(l, k)
\nonumber \\
& =
\Big\{
\big[{\mathbf{H}}(k) \mathbf{R}_\mathrm{X}  {\mathbf{H}}(k)^*\big]^{+}
\big[{\mathbf{H}}(k) \mathbf{I}_1\big] 
\Big\}^*
\mathbf{H}(k) \mathbf{X}(l, k)
\nonumber \\
& =  
X_\mathrm{s}^\mathrm{[2]}(l, k), \quad \forall l,k.
\label{eq:esti_proof}
\end{align}

\end{appendices}

\bibliography{references}
\bibliographystyle{ieeetr}

\end{document}